\documentclass[prd,letter,eqsecnum,nofootinbib,preprint,floatfix]{revtex4}
\usepackage{colordvi}
\usepackage{graphicx}
\usepackage{bm}
\usepackage{multirow}
\usepackage{enumerate}
\reversemarginpar
\def\<{\langle}
\def\>{\rangle}
\newcommand{\text}{\rm}
\def\Tr{{\rm Tr}\,}
\def\tr{{\text tr}\,}
\def\Eq#1{Eq.~(\ref{#1})}

\renewcommand\Re{{\text Re}\,}

\def\Sec#1{Section~\ref{#1}}
\def\Ref#1{Ref.~\cite{#1}}

\begin{document}

\begin{titlepage}

\mbox{} \hfill CAS-KITPC/ITP-137~~\\

\vspace*{0.7in}
 
\begin{center}
{\large\bf Large-\boldmath{$N$} spacetime reduction \\ and the sign and silver-blaze  problems  of dense  QCD}
\vspace*{1.0in}

\vspace{-1cm}
{Barak Bringoltz\\
\vspace*{.2in}
Department of Physics, University of Washington, Seattle,
WA 98195-1560, USA\\
and\\
Kavli Institute for Theoretical Physics China, CAS, Beijing 100190, China\\
}
\vspace*{0.25in}

\end{center}

\vspace*{0.15in}

We study the spacetime-reduced (Eguchi-Kawai) version of large-$N$ QCD  with nonzero chemical potential. We explore a method to suppress the sign fluctuations of the Dirac determinant in the hadronic phase; the method employs a re-summation of gauge configurations that are related to each other by center transformations. We numerically test this method in  two dimensions, and find that it successfully solves the silver-blaze problem. We analyze the system  further, and measure its free energy $F$, the average phase $\theta$ of its Dirac determinant, and its chiral condensate $\<\bar\psi\psi\>$. We show that $F$ and $\<\bar\psi\psi\>$ are independent of $\mu$ in the hadronic phase but that, as chiral perturbation theory predicts,  the quenched chiral condensate drops from its $\mu=0$ value when $\mu\sim ({\rm pion\,\,mass})/2$. Finally, we find that the distribution of $\theta$ qualitatively agrees with further, more  recent, predictions from chiral perturbation theory.

\end{titlepage}

\setcounter{page}{1}
\newpage
\pagestyle{plain}

\section{Introduction}
\label{intro}

It is of great  interest to calculate properties of four dimensional QCD at low temperatures and large, but not asymptotic, chemical potentials. Due to the `sign problem' that inflicts euclidean lattice Monte-Carlo simulations such calculations seem currently unrealistic \cite{PDF} (for recent progress in this see \Ref{PDF1}). Specifically, the fluctuations in the phase $\theta$ of the Dirac operator's determinant, which become particularly strong once the chemical potential $\mu$ grows beyond half the pion mass $m_\pi/2$, cause the simulations to fail.  At a first glance this seems surprising because, at least for low temperatures $T\ll \Lambda_{\rm QCD}$, one expects physics to depend on $\mu$ only once $\mu \stackrel{>}{_\sim} m_B/3$, with $m_B$ the lightest baryon mass. Thus, for $m_\pi/2 \le \mu \le m_B/3$, physical observables are approximately independent of $\mu$ but $\theta$ is strongly oscillating. The apparent tension between these two facts was termed the `silver-blaze' problem \cite{SB} and for small $\mu$ it can be studied with  chiral perturbation theory \cite{LSV}. The sign and silver blaze problems were also recently studied in scalar field theory using complex Langevin dynamics \cite{Gert} and in fermionic theories with four-fermi interactions \cite{SC09}.

In this paper we analyze various aspects of the QCD sign problem.
In particular, we ask how the sign problem behaves once we 
project the Dirac determinant to be neutral with respect to the $Z_N$ center symmetry of the $SU(N)$ group. Our current paper is not the first to discuss this `$Z_N$-averaging' --- see for example Refs.~\cite{early} --- but we are not aware of studies that explicitly checked its effect on the sign fluctuations of the determinant in the confined phase of QCD.

Our modest computational resources lead us to approach the problem in its large-$N$ `t Hooft limit and to utilize the space-time reduction of the planar theory. This reduction in degrees of freedom was discovered by Eguchi and Kawai in the seminal \Ref{EK} whose content is the following: under certain conditions, infinite-volume lattice large-$N$ QCD is  equivalent, on all distance scales, and in certain sectors of its spectrum, to its `volume-reduced' version. The latter is nothing but lattice QCD defined on a single lattice site. 

 The simplest implementation of the Eguchi-Kawai (EK) equivalence works only in two euclidean dimensions \cite{BHN,MK}, while in four dimensions, some extensions of the original EK prescription, like those of \Ref{AEK,DEK}, are expected to be successful. The numerical work involved in using  these extended prescriptions is comparably  demanding, 
 and their applicability is still being tested  numerically \cite{BS,HN}. These facts lead us to focus on the two-dimensional system in this paper. Provided that the constructions of Refs.~\cite{DEK,AEK} survive their numerical tests, we do not see any obvious numerical or conceptual difficulties in extending the present approach to four dimensions.

Below is the outline of our paper. We begin in \Sec{restrict} by reminding the reader what are the validity conditions of the EK equivalence and discuss what they imply for our numerical calculations. We then present the construction of the two-dimensional EK theory in \Sec{def_EK}, and discuss how we simulate it and which observables we measure. We formulate $Z_N$-averaging in \Sec{ZN}. Then, in \Sec{ZN_works?}, we discuss the connection of the sign and silver-blaze problems to the signal to noise ratio (SNR) of physical observables calculated with  a Monte-Carlo simulation, and the way $Z_N$ averaging can potentially solve these  problems. In \Sec{Sp} we present our numerical study of the sign problem and show the average sign of the determinant as a function of $\mu$. The tests of $Z_N$-averaging are presented in Section~\ref{ZN_test}. We then show results from measurements of various physical observables in \Sec{physical}, and in \Sec{phasedetD} we present the distributions of $\theta$ as a function of $\mu$, and the way the eigenvalues of the Dirac operator scatter in the complex plane for different values of $\mu$.
In \Sec{summary} we summarize our study.

\section{Eguchi-Kawai space-time reduction at nonzero density}
\label{restrict}

There are two issues one should be aware of when one studies nonzero density with space-time reduction.  We discuss these issues in Sections~\ref{EK_val} and \ref{EK_sat}, where we show that the method of EK space-time reduction can be used only within the hadronic phase. Despite that, this method is useful to understand the silver blaze and sign problems of QCD at nonzero $\mu$, and we emphasize this point  in \Sec{imply}.

\subsection{The importance of translation symmetry}
\label{EK_val}

In \Ref{axial} we studied the space-time reduction of large-$N$ QCD in the presence of baryons. To have full analytic control we focused there on the two-dimensional version of QCD -- the `t Hooft model. We found results which are consistent with the general validity conditions that non-perturbative large-$N$ equivalences require. Specifically, we saw that the main validity condition for volume reduction (or the `Eguchi-Kawai equivalence') in our context is that translation symmetry is intact in the infinite-volume  field theory.\footnote{Another condition is that the $Z_N$ center symmetry of the theory is intact, but  this is dictated by the leading gluon dynamics and not by the fermions.}
 To understand this recall that the EK theory is the single-site version of QCD. Thus, it is a projection of QCD that removes degrees of freedom with nonzero momentum. As such, it is clear that using the EK equivalence does not make sense when the vacuum of QCD spontaneously breaks translation invariance and is characterized by condensates which carry nonzero momentum.  The question of whether translations are a symmetry of the QCD ground state at nonzero $\mu$ is a dynamical one, and in \Ref{nonzeroB} we showed that in two dimensions, for any quark mass, and for a single flavor, this symmetry breaks spontaneously for $\mu>m_B/N$ (for earlier results concerning only the chiral limit see \Ref{SchonThies}). Simple arguments lead one to expect a similar phenomenon in higher dimensions (see discussion in section IV of \Ref{nonzeroB}).
This means that the single site theory that we study in the current paper is equivalent to the infinite-volume field theory throughout the hadronic phase which, in two dimensions and in the single flavor case, is bounded by $\mu\le m_B/N$. Beyond that point the theory we study here can be thought of as a complicated matrix model. 

\subsection{Implications of lattice saturation}
\label{EK_sat}
 
As any lattice field theory, volume-reduced QCD can be expected to be influenced by lattice artifacts at sufficiently large values of $\mu$. In particular, for lattice spacing $a$ and chemical potential $\mu \sim O(1/a)$, the density is at the cutoff scale, and the lattice is saturated with baryons -- there is an $O(1)$ number of baryons per site. This regime, in which the Pauli exclusion on each site is saturated or nearly saturated,  is usually referred to as the lattice saturation regime and is governed by lattice artifacts.

 What does the saturation phenomenon imply on a single-site theory? In $d+1$ dimensions the spatial volume $V$ of the single-site theory is $a^d$, and as long as $\mu$ is smaller than a critical value $\mu_c$, there are no baryons in the vacuum (as mentioned above, for $d=1$ and with a single flavor, baryons repel and $\mu_c=m_B/N$ \cite{SNL,nonzeroB}). But when $\mu>\mu_c$ the hadronic phase makes way to a phase that accommodates baryons, and the baryon number $B$ is then at least $1$. Importantly, this makes the baryon density, $B/V$, of $O(1/a^d)$ and so at the cutoff scale. 

Therefore, we see that we cannot accommodate physical {\em nonzero} densities on a single site: when the baryon number is nonzero, it is at least one, the density of the single-site theory is at the cutoff scale.

\subsection{Usefulness of space-time reduction at nonzero {\boldmath $\mu$}.}
\label{imply}

At this point an obvious question comes to mind --- what is the usefulness of the EK single-site theory at $\mu>0$ and why do we wish to study it in this paper? The answer is that we wish to understand the sign and silver blaze problem of the hadronic phase. In that phase $\mu$ is nonzero, but the density {\em is} zero and so we can use the EK theory to analyze the theory there.

Also, despite the fact that in the saturation regime (outside the hadronic phase), the EK theory is not equivalent to the field theory, we can still ask how the EK model behaves there, and see whether we can learn something about the general properties of the saturation regime (as we alluded to above, this regime appears also in the standard, infinite-volume, `non-reduced' lattice field theory). 

\section{Lattice details of the Eguchi-Kawai theory at nonzero {\boldmath $\mu$}: the action and the simulation algorithm}
\label{def_EK}

The EK theory or `volume-reduced QCD' is a lattice gauge theory defined on a single lattice site.  Since our numerical efforts are focused on two dimensions, we define below the construction of the EK theory only in this case. 

\subsection{Definition of path integral}
\label{def_PI}

The path integral is 
\begin{eqnarray}
Z &=& \int DU \, \int D\psi \, D \bar{\psi} \, \exp\left(S_{\rm YM} + S_F \right),\label{Z}\\
S_{\rm YM} &=& 2Nb \,\, \Re \, \Tr\, \left(U_1 \, U_2\, U^\dag_1\, U^\dag_2\right),\label{SYM1}\\
S_{\rm Dirac} &=& \bar{\psi} \, D\, \psi,\\
D &=&  \hat m + \frac12 \gamma_1 \left(U_1\, e^{\hat \mu }-e^{-\hat \mu }\, U^\dag_1\right) + \frac12 \gamma_2 \left(U_2-U^\dag_2 \right),\label{D}
\end{eqnarray}
Here $U_{1,2}$ are $SU(N)$ matrices and $DU=dU_1 dU_2$ with $dU_{1,2}$,  the Haar measure on $SU(N)$. $\psi$ is a two dimensional Dirac spinor of the `naive' fermion type and it transforms in the fundamental representation of the gauge group. Thus, it corresponds to $4\times N_f$ Dirac fermions in the continuum. The $2N_fN\times 2N_fN$ matrix $D$ is the euclidean Dirac operator of a lattice gauge theory on a single site. The quark mass $m$ and chemical potentials $\mu$ are related to their dimensionless lattice quantities $\hat m$ and $\hat\mu$ in the usual way by the lattice spacing $a$. 
In two dimensions the gauge coupling $g$ has dimensions of mass and the standard dimensionless lattice coupling $\beta$ is defined by 
\begin{equation}
\beta=\frac{2N}{ a^2g^2}.
\end{equation}
In the `t Hooft limit $g^2$ scales like $O(N^{-1})$ and so it is useful to define a new lattice coupling $b=\beta/(2N^2)$ that scales like $O(N^0)$ in the large-$N$ limit -- this is the coupling that appears in \Eq{SYM1}. In terms of the dimensionful `t Hooft coupling $\lambda=g^2N$, the coupling $b$ is given by 
\begin{equation}
b=\frac1{ a^2\lambda}.
\end{equation}
All dimensionful quantities like the quark mass $m$ and the baryon chemical potential $\mu$ will be fixed in units of $\lambda$ when we take the continuum limit $b\to \infty$. Also, to make a connection with \Ref{KNN}, we define the parameter $\gamma$ that controls the quark mass 
\begin{equation}
\gamma = \pi \frac{m^2}{\lambda}.\label{gamma}
\end{equation}
In terms of $b$ and $\gamma$ the lattice quantities are given by
\begin{eqnarray}
\hat m &\equiv &am= \frac{m}{\sqrt{\lambda}} \, \frac1{\sqrt{b}} = \sqrt{\frac{\gamma}{\pi b}},\\
\hat \mu &\equiv &a\mu= \frac{\mu}{\sqrt{\lambda}} \, \frac1{\sqrt{b}}.
\end{eqnarray}

It is useful to note the following results on the spectrum of the `t Hooft model in the large-$N$ limit: The pion mass at zero chemical potential for small and large values of $\gamma$ is given by (see  \Ref{KNN})
\begin{equation}
\frac{m_\pi}{\sqrt{\lambda}} \simeq  \left\{
\begin{array}{lr}
1.08 \, \gamma^{1/4}  + O(\gamma^{1/2}) & \quad \gamma \ll 1,\\
1.13 \, \gamma^{1/2}  + O(\gamma^{-2/3}) &\quad  \gamma \gg 1, 
\end{array}
\right. \label{m_pi}
\end{equation}
which is expected to hold for any number of flavors $N_f$ that obeys $N_f\ll N$.
Next, the lightest excitation with a nonzero baryon number {\em in the single-flavor case} is the baryon  whose mass in the chiral limit is \cite{SNL}
\begin{equation}
\frac{m_B}{N} = \frac{4}{\pi}\times\frac{m_\pi}2.\label{mBChi}
\end{equation}
Indeed, in two dimensions, both $m_\pi$ and $m_B$ vanish for massless quarks! For a non-zero quark mass, however, $m_B/N>m_\pi/2$ and we have the intermediate range, $m_\pi/2\stackrel{<}{_\sim}\mu\stackrel{<}{_\sim} m_B/N$, where the theory has the same properties as those at $\mu=0$, but where the sign fluctuations are strong --- just like in the case of physical four dimensional QCD.

The most recent calculation of the $m_B/N$ in $1+1$ dimensions (for $N_f=1$), away from the chiral limit, was done in \Ref{nonzeroB}. The previous study in \Ref{SNL} had certain ingredients missing, whose effect was calculated in \Ref{nonzeroB}, and was found to be small. In any event, using the methods of \Ref{nonzeroB}, we evaluated the baryon mass for the quark mass studied  in the current paper and we find\footnote{This estimate was obtained in the Hamiltonian formalism at a spatial lattice spacing of $a\sqrt{\lambda}\simeq 0.31$. The lattice spacing corrections are expected to be small for this value of $a$ \cite{nonzeroB}.}
\begin{equation}
\frac{m_B}{N\sqrt{\lambda}} \simeq 0.758 \quad {\rm at} \quad \gamma=1.\label{mB}
\end{equation} 
We note that both the estimates for $m_B$ in Eqs.~(\ref{mB}) and~(\ref{mBChi}) are for $N_f=1$ and we are not aware of any extensions to general $N_f$. Since we define the EK theory with  naive fermions, for sufficiently small lattice spacings we actually have four physical Dirac fermions. In the absence of any estimates for the baryon mass in the four-flavor case, we proceed by  assuming that the $N_f$ dependence of $m_B$ is weak.

\subsection{Calculation strategy}
\label{strategy}

Expectation values are measured with  Monte-Carlo simulations that use the Yang-Mills action to generate gauge configurations. This means that
 $\<{\cal O}\>$ is approximated by the sum over gauge configurations that represent the YM ensemble:
\begin{equation}
\<{\cal O}\>\simeq \frac1{M}\, \sum_{{\cal C}=1}^{M} \, {\cal O}_{\cal C}.
\end{equation}
Here $M$ is the number of configurations. Our measurements can be therefore referred to as performed via `reweighting' because we generate the gauge configurations with the YM action, and reweight in the fermion determinant. 

This reweighing would not be necessary had we focused on $\mu=0$ since the large-$N$ theory is naturally quenched there (at least at nonzero quark mass); in practice this would be reflected in a Monte-Carlo by the fact that the `fermionic force' in a Hybrid Monte-Carlo is $O(1/N)$ suppressed compared to the gluonic force. At $\mu>0$, however, reweighting in the phase of the determinant is crucial (as our results would show).  

By reweighting from the YM ensemble, we use the fact that the important  gluonic configurations at large-$N$ are chosen by the YM action, and that the fermions do not back-react to the gluons. This absence of back-reaction is characteristic of the large-$N$ limit and  can be justified in various approaches (like Hamiltonian coherent states) but we will not discuss this issue here (see instead the forthcoming \Ref{BY}). We should note that the lack of back-reaction at large-$N$ persists at nonzero $\mu$, but that this does not mean that the quenched approximation should work there (despite the common lore, the quenched approximation is {\em not} equivalent to removing the back-reaction of quarks on gluons from QCD -- more details will be given in \Ref{BY}).

In the context of standard lattice QCD studies, the reweighting procedure is known to be highly susceptible to significant systematic errors when $\mu$ and the lattice volume are large enough \cite{PDF}. These errors can be diminished, however, if one scales the number of sampled field configurations {\em exponentially} in the size of the matrix $D$. In four dimensional $3$-color QCD one has ${\rm dim}(D) = 12 \times \, ({\rm lattice \,volume})$, and so going to the thermodynamic limit is not realistic with current computational power. Similarly, in our case we have ${\rm dim}(D)=2\times N$ and our computational resources allow us to study only  $N\le 60$.\footnote{Our errors modestly increase with $N$, and this is because we do not actually increase our statistical sample exponentially. Nevertheless, we find that $N=40$ and $60$ are already close enough to the large-$N$ limit and in most cases have errors that are sufficiently small for our purposes. Indeed, this is how large-$N$ reduction plays an important role in our study.}

To simulate $S_{\rm YM}$ we used the heat-bath algorithm introduced in \cite{FH}. Measurements of fermionic quantities are typically separated by $100-1000$ full updates of  the model,\footnote{By `full update' we mean updating all the $N(N-1)/2$ $SU(2)$ subgroups of each of the two matrices $U_{1,2}$.} and thus are expected to be uncorrelated. To check this we estimate errors with a jackknife procedure.  
Below we list the fermionic quantities that we measure for most values of $b$, $N$ and $\mu$.

\renewcommand{\labelenumi}{\alph{enumi}.}

\begin{enumerate}

\item The fermionic contribution to the free energy, $F$. This was obtained from 
\begin{equation}
F = \frac1{N} \log Z_{QCD} \equiv  \frac1{N}\,\log \left\< \det \, D(\mu) \right\>. \label{F1}
\end{equation}

\item The average sign of the determinant given by 
\begin{equation}
\<\cos(\theta)\> \equiv \left\< \Re \left( \frac{\det \, D(\mu)}{\left|\det \, D(\mu)\right|}\right) \right\>. \label{Sign}
\end{equation}

\item The quenched quark condensate which we define to be
\begin{equation}
\<\bar{\psi} \psi \>_{\rm quenched}  =\left\< \tr \left( D^{-1}(\mu) \right)\right\>. \label{SigmaQ}
\end{equation}

\item The physical unquenched quark condensate 
\begin{equation}
\<\bar{\psi} \psi \>_{\rm unquenched}  =\frac{\left\< \tr \left( D^{-1}(\mu) \right)\times \det D\right\>}{\<\det D\>}. \label{Sigma}
\end{equation}

\item The distribution of the angles $\theta$ and $\alpha$ defined via
\begin{eqnarray}
\det D &=& \left|\det D\right|\, e^{i\theta}, \label{theta}\\
\bar\psi\psi &=& \left|\bar\psi\psi\right|\, e^{i\alpha}.\label{alpha}
\end{eqnarray}

\item The way the eigenvalues of $D$ scatter in the complex plane.

\end{enumerate}

Finally, for $\mu=0$ and for each value of $b$ and $N$ we also measured the pion propagator of momentum $q$ along the ``$2$'' direction using the Gross-Kitazawa momentum injection method \cite{GK} (that was also used in \Ref{GK_use}). This means that we measure 
\begin{equation}
G_{\pi}(q) = \left\< \tr \, \left(D^{-1}(U_2 e^{iq/2}) \, \gamma_5 \, D^{-1}(U_2 e^{-iq/2}) \gamma_5\right) \right\>,
\end{equation}
and extract the pion mass by measuring the effective mass $m_{\rm eff}$ defined by
\begin{equation} 
m_{\rm eff} = -\lim_{x\to \infty} \, \frac{\partial \left[ \log \, \left( \Re \int \, dq \, e^{iq\, x} G_\pi(q) \right)\right]}{\partial x}.  \label{meff}
\end{equation}

We calculated these observables for chemical potentials in the range $\mu/\sqrt{\lambda}\in[0,3]$. Throughout our calculation we fixed $\gamma=1$ and so we expect $m_\pi/\sqrt{\lambda}\simeq 1.2$ from \Eq{m_pi} and $m_\pi/\sqrt{\lambda}\simeq 1.5$ from the continuum extrapolations of the numerical results in \cite{KNN}. We perform lattice simulations at $b=0.6,6.0,10.0$. The number of field configurations that we used for each choice of $N$ and $b$ is given in Table~\ref{lat_parm}. We also list the pion mass in each case as measured from \Eq{meff}. For brevity, we set $\frac12 m_\pi/\surd\lambda = 0.65^{+0.03}_{-0.05}$ to encompass all the values of $m_\pi/2$ from Table~\ref{lat_parm}.  We present the results of our study in the next sections. 

The numerical routines we used were defined with double precision, and to check whether our calculations are sensitive to this (especially our evaluations of the determinants) we recalculated the free energies and average signs with quadruple precisions and 
confirmed that our results remained the same.
\begin{table}[htb]
\setlength{\tabcolsep}{4mm}
\begin{tabular}{cccc}
\hline\hline
$N$ & $b$ & no. of configurations & $m_\pi/\sqrt{\lambda}$\\ \hline 
\hline\hline
 & $0.6$ & $16000$ & $\stackrel{<}{_\sim} 1.20$ \\
$10$ & $6.0$ & $20000$ & $1.24(12)$\\
 & $10.0$ & $20000$ & $1.24(13)$\\ \hline
 & $0.6$ & $12000$ & $\stackrel{<}{_\sim} 1.24$\\
$20$ & $6.0$ & $120000$& $1.25(5)$\\
 & $10.0$ & $120000$ & $1.33(6)$\\ \hline
 & $0.6$ & $1000$& $\stackrel{<}{_\sim} 1.29$\\
$40$ & $6.0$ & $105000$& $\stackrel{<}{_\sim} 1.23$\\
 & $10.0$ & $101000$ & $\stackrel{<}{_\sim} 1.40$\\ \hline
$60$ & $10.0$ & $50000$& -- \\ \hline
\end{tabular}
\caption{Details of runs used to map the phase diagram in $\mu$. The number of full model updates was of $O(10^6-10^7)$ depending on the value of $N$. Not all configurations were used to estimate the meson masses. The cases in which we only give an upper bound on $m_\pi$ are those in which we did not observe the asymptotic mass plateau in the pion propagator. The reason for this is either that $m_\pi$ was too large in lattice units, and so falls into the noise quickly (we saw this for $b=0.60$), or that one needs to calculate the pion propagator at separations $x$ which are too large for our resources to accommodate.
}
\label{lat_parm}
\end{table}

\section{Definition of $Z_N$-averaging}
\label{ZN}

In this section we define $Z_N$-averaging and briefly explain its logic. This averaging relies on the $Z_N$ symmetry of the YM part of the action and of the Haar measure, and it is defined by the following prescription. 

For each gauge configuration ${\cal C}$, characterized by the gauge fields $(U_1,U_2)$, perform the replacement
\begin{equation}
{\cal O}_{\cal C}\, (U_1,U_2) \to \left({\cal O}_{\cal C}\right)_Z\equiv \frac1{N^2} \sum_{k_1,k_2=1}^{N} {\cal O}_{\cal C}\left(U_1\, e^{2\pi i k_1/N}, U_2\, e^{2\pi i k_2/N}\right),\label{ZN_def}
\end{equation}
and only then average over the field configurations $U_{1,2}$. We distinguish between the results obtained with and without this $Z_N$-averaging by adding the subscript $Z$ to the former, when necessary.  In four dimensions \Eq{ZN_def} is generalized to have four sums over $k_{1,2,3,4}\in [1,N]$.

Let us explain the logic behind this proposal. We expect that the partition function will depend on $\mu$ in a very special way; for a $U(N)$ group there should be no $\mu$-dependence since there are no gauge invariant charges that couple to $\mu$. For an $SU(N)$ group, however, there {\em can} be $\mu$-dependence, but only through the combination $\mu N$ since the only gauge invariant excitations that couple to $\mu$ are baryons, and these have charges that are integer multiples of $N$. 

How can we anticipate these results from the path integral in \Eq{Z}? Expand $\det D$ in \Eq{Z} in the worldline approach \cite{worldline}. It is then easy to see that the $\mu$-dependence we discuss above arises because both the measure of the path integral and the YM action are symmetric under the center of the gauge group. Specifically, the $\mu$ dependence of the determinant comes in through terms of the form
\begin{equation}
e^{q\mu}\times  \prod_i \, \left(\tr U^{k_i}_1\right)^{p_i}, \quad {\rm and}\quad e^{-q\mu}\times \left[\prod_i \, \left(\tr U^{k'_i}_1\right)^{p'_i}\right]^\star,\label{terms}
\end{equation}
with $q=\sum_i k_ip_i=\sum_i k'_i p'_i$, but  the center symmetry allows $q=0$ for $U(N)$ and $q/N=$ integer for $SU(N)$; $Z_N$-averaging enforces the center symmetry on each gauge configuration, and automatically makes the path integral depend on $\mu$ as prescribed above. In fact, since it is only $U_1$ that appears in \Eq{terms}, we can also try and do a partial $Z_N$ averaging by re-summing gauge configurations that differ by a center transformation only in the time direction. 
\begin{equation}
{\cal O}_{\cal C}\, (U_1,U_2) \to \left({\cal O}_{\cal C}\right)_{Z-{\rm partial}}\equiv \frac1{N} \sum_{k_1=1}^{N} {\cal O}_{\cal C}\left(U_1\, e^{2\pi i k_1/N}, U_2\right).
\end{equation}
In our two-dimensional case this saves a factor of $N$ in the measurements of observables (and $N^{d-1}$ for $d$ space-time dimensions). Numerically we see that, at least in the hadronic phase, such a partial averaging is almost as good as the full $Z_N$ averaging (see \Sec{ZN_test}), and so we use it when our resources are insufficient to perform the latter (for example when we calculate the unquenched quark condensates for $SU(40)$ or the average sign for  $SU(60)$).

\section{Can $Z_N$ averaging solve the silver-blaze and sign problems?}
\label{ZN_works?}

As explained above, one reason that we expect $Z_N$ averaging to be useful is that it makes the anticipated dependence of the partition function $Z_{\rm QCD}$ on $\mu$ manifest on a configuration-by-configuration basis --- it is a noise reduction technique. To make this expectation more precise we need to discuss the source of the statistical noise in the calculation of $Z_{\rm QCD}$ for $\mu>0$ (and its relation to the sign problem).\footnote{For a  previous useful discussion on  see \Ref{Kratochvila}.} We do so in \Sec{SNR}, and then move on to  \Sec{2scenarios} where we discuss how such noise can be reduced when one uses $Z_N$ averaging.

\subsection{The sign problem, the signal to noise ratio, and pion physics.}
\label{SNR}

The sign problem and the signal to noise ration (SNR) of observables in QCD are closely related. To show this, let us discuss the SNR of observables in a Monte-Carlo that generates gauge configurations for the `Phase-Quenched' (PQ) ensemble. The measure of this ensemble is proportional to the absolute value of the Dirac determinant.
Specifically, if we write 
\begin{equation}
\det D= \left| \det D \right| e^{i\theta},
\end{equation}
then, when calculating the expectation value of 
an observable ${\cal O}$ with gauge configurations of the PQ ensemble, one needs to treat the phase $e^{i\theta}$ as part of the observable. This means that the QCD average of ${\cal O}$ becomes
\begin{equation}
\<{\cal O}\>_{\rm QCD}\equiv \frac{\<{\cal O} e^{i\theta}\>_{\rm PQ}}{\<e^{i\theta}\>_{\rm PQ}} \label{PQ}.
\end{equation}
It is useful to note that we can define the PQ average $\<,\>_{\rm PQ}$ through the Yang-Mills average $\<,\>_{\rm YM}$ 
\begin{equation}
\<{\cal O}\>_{\rm PQ}\equiv \frac{\<{\cal O} |\det D|\>_{\rm YM}}{\<|\det D|\>_{\rm YM}}.
\end{equation}

From \Eq{PQ} we see that the SNR of any observable gets a contribution from the SNR of $\<e^{i\theta}\>_{\rm PQ}=\<\cos \theta\>_{\rm PQ}$ (here we assume charge conjugation). The SNR of $\<\cos \theta\>_{\rm PQ}$ is given by
\begin{equation}
{\rm SNR}^{\rm PQ}_{\cos\theta}\equiv \frac{\left(\<\cos \theta\>_{\rm PQ}\right)^2}{\<\cos^2\theta\>_{\rm PQ}}. \label{SNR4}
\end{equation}
In terms of $\<,\>_{\rm YM}$, \Eq{SNR4} becomes
\begin{equation}
{\rm SNR}^{\rm PQ}_{\cos\theta}\equiv 
\frac
{\left( \<\cos \theta |\det D|\>_{\rm YM}\right)^2}
{\<\cos^2\theta |\det D| \>_{\rm YM}\<|\det D|\>_{\rm YM}}=\frac{\left( \<\det D\>_{\rm YM}\right)^2}{\<\cos^2\theta |\det D| \>_{\rm YM}\<|\det D|\>_{\rm YM}},\label{SNRPQ}
\end{equation}
and it is now clear that if there is no sign problem, and $\cos\theta\approx 1$ for the majority of field configurations, then ${\rm SNR}^{\rm PQ}_{\cos\theta} \simeq 1$. 


It is instructive to  show how pion physics tends to appear when there is a sign problem. This can be seen by assuming that a sign problem makes $\cos ^2 \theta \simeq \frac12$ and turns \Eq{SNRPQ} into
 \begin{equation}
{\rm SNR}^{\rm PQ}_{\cos\theta} \simeq 2 \left( \frac{\<\det D\>_{\rm YM}}{\<|\det D|\>_{\rm YM}}\right)^2.\label{SNR2}
\end{equation}
Now, for simplicity,  consider the case of two degenerate flavors and note that 
\begin{eqnarray}
{\rm SNR}^{\rm PQ}_{\cos\theta} \simeq 2 \left( \frac{\<\det D^2_1\>_{\rm YM}}{\<|\det D_1|^2\>_{\rm YM}}\right)^2 = 2 \, e^{+2V/T \left( f_{I} - f_B \right)}.\label{SNR2.5}
\end{eqnarray}
Here $D_1$ is the Dirac operator of a single flavor and $f_{B}$ and $f_I$ are the free energy densities of two-flavor QCD in the presence of baryon number and isospin chemical potentials that are both equal to $\mu$. To proceed, note that baryons do not contribute to $f_B$ for $\mu \lesssim m_B/N$ (we always restrict to low temperatures) which gives
\begin{equation}
f_B(\mu) = f_B(0)\quad {\rm for}\quad \mu<m_B/N.
\end{equation}
In contrast, $f_I$ changes when $\mu>m_\pi/2$  (the iso-spin system goes through pion condensation above this value of $\mu$). These facts, together with $f_I \le f_B$, which is true because $\<\left|\det D^2_1\right|\>_{\rm YM} \ge \<\det D^2_1\>_{\rm YM}$, tell us that ${\rm SNR}^{\rm PQ}_{\cos \theta}$ is exponentially small in  the four-volume $V/T$ when 
\begin{equation}
m_\pi/2\le \mu\lesssim m_B/N.\label{problem1}
\end{equation}

\subsection{Two scenarios for the numerical cause of the sign problem and how {\boldmath $Z_N$} averaging can solve it}
\label{2scenarios}

We can understand the problem of the exponential suppression of ${\rm SNR}^{\rm PQ}_{\cos\theta}$ and the associated sign problem from the point of view of the worldline approach. In particular, these problems reflect sign fluctuations in $\det D$ that are caused by the contribution to $(\det D)$ from Polyakov loops that wrap around the temporal direction. From this point of view, there  are two scenarios for the cause of the sign problem:

\subsubsection{Scenario no.~1: The sign problem is caused by fluctuations of Polyakov loops that wrap the torus $k$ times with $k/N\neq$ integer.}
\label{s1}

Polyakov loops whose winding number around the temporal direction $k$ is not an integer multiple of $3$ (or $N$ in $SU(N)$) are unphysical; by the center symmetry these nonzero $N$-ality worldlines have zero expectation value and so they do not contribute to the numerator of \Eq{SNR2.5}. This is despite the fact that they appear in $\det D^2_1$ with enhancing fugacity factors of $e^{k\mu/T}$.

While the unphysical worldlines do not contribute to $\<\det D^2_1\>_{\rm YM}$, they do contribute to its noise $\<\left|\det D_1 \right|^2\>$ appearing in the denominator of \Eq{SNR2.5}; despite the fact that their magnitude can be small, their fugacity factors $e^{k\mu/T}$ can be large, and they can cause significant fluctuations in the overall fluctuating phase of the determinant. The magnitude squared of a $k$-worldline contribution to $\det D^2_1$ is a worldline configuration with $k$ worldline-antiworldline pairs which can be interpreted as a worldline of $k$ pions (this can be seen explicitly from the worldline expansion of the $\left|\det D_1\right|^2$). Therefore, this magnitude squared can be estimated by $e^{-k m_\pi/T}$, and we can now 
anticipate that $\det D^2_1$ will have significant noise when $e^{k\mu/T} \times e^{-k m_\pi/2T}\gg 1$ or when $\mu > m_\pi/2$, as we saw in \Eq{problem1}.\footnote{In principle we should replace $km_\pi$ by the free energy of a system with isospin number equal to $k$, but the binding energy of $k$ pions is suppressed at large-$N$.}

If the scenario we describe above is the one causing the sign problem then it is an optimistic scenario: worldlines that wrap the euclidean time torus $k$ times with $k/N\neq$ integer can be removed from the noise on a configuration-by-configuration basis using $Z_N$ averaging. This would solve the silver-blaze problem since it would remove the phase fluctuations of the Dirac determinant in the hadronic phase. 

Importantly, however, our focus on the unphysical worldlines in the current scenario assumes that the physical, `baryonic', worldlines with $k/N=$ integer, are suppressed on a configuration-by-configuration basis (assuming, for example, that they are weighted by $e^{-km_B/T}$). If that is the case then the only sign fluctuations that would survive $Z_N$ averaging would occur when the baryonic worldlines contribute to $\<\det D^2_1\>$, i.e.~when $\mu > m_B/N$. In the next scenario below we ask what happens if this is {\em not} the case.

\subsubsection{Scenario no.~2: The sign problem is caused by Polyakov loops that also have $k/N=$ integer.}
\label{s2}

Unfortunately, the physical, `baryonic', worldlines, for which $k$ is an integer multiple of $N$ (and therefore are insensitive to $Z_N$ averaging), can have a sizable contribution to the sign and SNR problems even at small $\mu$. Naively one might think that these worldlines are suppressed so long as $\mu \lesssim m_B/N$ (see end of previous section) but this is not necessarily correct. To see why, let us  focus on the single-baryon contribution to $(\det D)$, and denote it by $\left(\det D\right)_{B=1}$. The latter reflects the contributions of all worldlines that wrap the euclidean circle $N$ times and so it comes with a fugacity pre-factor of $e^{N\mu/T}$.  Our main point here is that there is no a priori reason why these contributions should be suppressed by $e^{-m_B/T}$ on a configuration-by-configuration basis. Quite the contrary, it is possible that there are some gauge configurations where these contribution 
are of $O\left(e^{-N m_\pi/2}\right)$  --- here we view the weight of a single quark line as the weight of half of a pion.\footnote{Indeed, the noise in the evaluation of $\left(\det D\right)_{B=1}$ is given by the magnitude squared of $\left(\det D\right)_{B=1}$, and can describe either a baryon-antibaryon pair or an $N$-pion system (both can be described by $N$ worldline-antiworldlines pairs).} 

Thus, we propose that $\left(\det D\right)_{B=1}$ evaluated on a single gauge configuration would have the following generic form reflecting the fact that its noise (or absolute value squared) has a combined origin of both a baryon-antibaryon pair or $N$ pions.
\begin{equation}
\left(\det D\right)_{B=1} \approx e^{\left(N\mu-m_B\right)/T} + c\times e^{N\left(\mu-m_\pi/2\right)/T}.\label{det1}
\end{equation}
Here the (generally complex) number $c$ depends on the dynamical details of the theory and on a configuration-by-configuration basis can be of an $O(1)$ magnitude but strongly fluctuating. In particular, the fluctuations in $c$ can average the second term in \Eq{det1} to zero, such that the gauge-configuration average will make $\<\left( \det D\right)_{B=1}\>$ (that determines the baryon mass) associated only with the $N$ world-lines that bind into a baryon.

From \Eq{det1} we see that if we send $T\to 0$, then the second term is the dominant one (since $m_\pi$ is always smaller than $2m_B/N$). This means that when $\mu > m_\pi/2$ pions will proliferate in the noise of $\det D$ and there will be a  severe sign problem.
In contrast, if we keep $T$ finite (but low), 
 and if  $|c|$ happens to be numerically very small, or more precisely if
\begin{equation}
|c| \ll e^{(N/2 m_\pi-m_B)/T},\label{whenZN}
\end{equation}
then the pionic term can become irrelevant. In that case $\left(\det D\right)_{B=1}$ will be of the order of $e^{-m_B/T}$ for most of the gauge configurations, and all the $\mu$ dependence of $\det D$ will be exponentially suppressed if $\mu<m_B/N$. Then $\det D$ will behave like it does at $\mu=0$, i.e.~there will be no sign problem. 

It is \Eq{whenZN} that determines whether $Z_N$ averaging solves the sign problem or not.
Similar arguments lead the search for optimized baryonic wave functions 
that couple minimally to pion physics, thus improving the signal to noise ratio of baryon correlation functions \cite{Detmold}. In that context, one searches for a `golden window' in the euclidean time separation of baryonic correlators, where their SNR is not exponentially small. Here, \Eq{whenZN} defines a golden window in the temperature $T$.

In the following sections we present results from numerical simulations, some of which done in order to determine whether $Z_N$ averaging indeed removes the sign fluctuations in the hadronic phase and solves the associated silver-blaze problem. Put differently we aim to see whether it is scenario no.~1 or no.~2 that takes place in our system.

\section{Results : the sign problem}
\label{Sp}

\begin{figure}[h!]
\centerline{
\includegraphics[width=15cm,height=10cm]{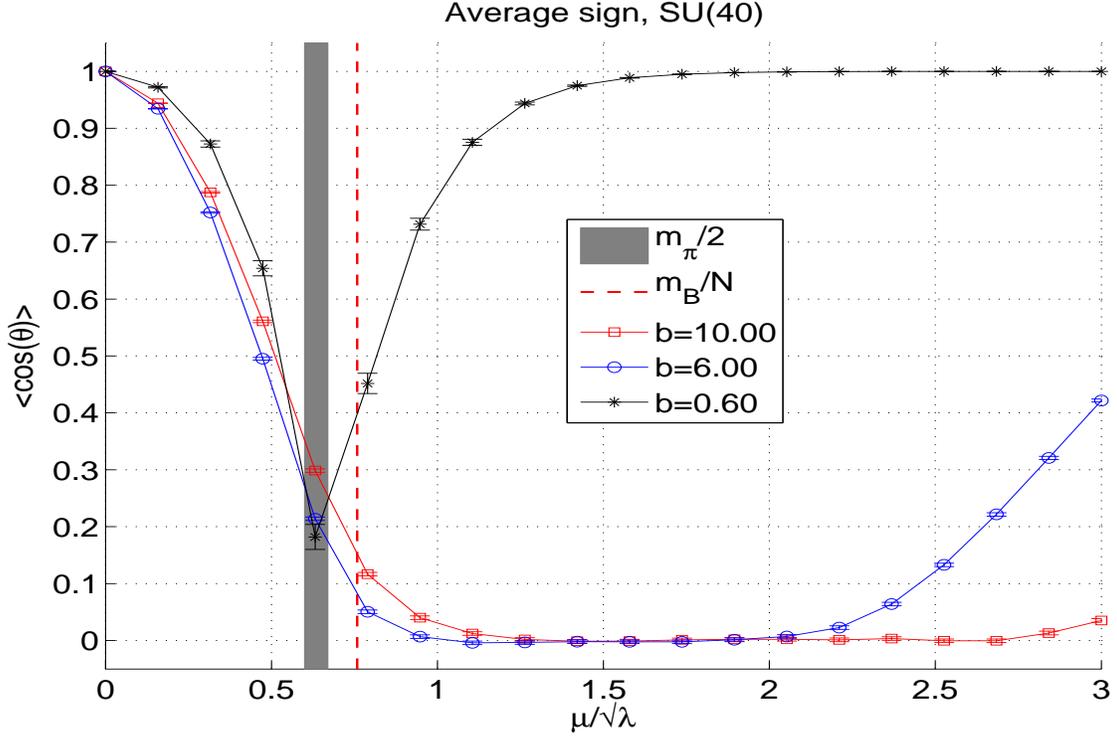}
}
\caption{The average sign, $\<\cos(\theta)\>$ for $SU(40)$ and $b=0.60,6.00,10.00$.}
\label{AvSign1}
\end{figure}

In this section we present the numerical analysis we performed for the averaged sign.
We begin by presenting the way the average sign of $\det D$ behaves as a function of $\mu$ for $SU(40)$ and different values of $b$ in Fig.~\ref{AvSign1}. 
The vertical (gray) solid band in the figure denotes our estimate for $m_\pi/2\approx 0.65 \surd\lambda$ (see above). The dashed vertical (red) line denotes the baryon mass $m_B$ (divided by $N$) for the single-flavor case. It is higher than $m_\pi/2$ by only about $16\%$ (this proximity is special to $1+1$ dimensions). As we emphasize in \Sec{restrict}, the EK theory that we study is equivalent to the $2d$ gauge theory on an infinite volume {\em only} below $m_B/N$. We nevertheless present here  (and below) the results for larger values of $\mu$, since for these values of $\mu$ the model has a saturation behavior which is accompanied by  an interesting behavior in the average sign. Specifically, from \Eq{D} we see that when the EK theory is saturated with baryons, it has no sign fluctuations because the $e^{\hat\mu}$ term governs the behavior of $D$ and one has
\begin{equation}
D \sim U_1 e^{\hat \mu}\left( 1 + O(e^{-\hat \mu})\right).
\end{equation}
Since $U_1\in SU(N)$ this means that 
\begin{equation}
\det D \stackrel{\rm saturation}{\longrightarrow} e^{2N\hat\mu},\label{sat}
\end{equation}
and that $\<\cos\theta\>\to 1$. Looking at Fig.~\ref{AvSign1}, we see that this is happening at $\mu_{\rm saturation}/\sqrt{\lambda}\simeq 1.5$ for $b=0.60$, and starting to happen at  $\mu_{\rm saturation}/\sqrt{\lambda}\simeq 3.0$ for $b=6.00$. The increase with $b$ of $\mu_{\rm saturation}$ is expected since in the continuum limit, $b\to \infty$, the saturation goes away completely (recall that this saturation reflects the fact that the baryon density on a lattice is bounded from above by the `saturation density', and the latter is of $O(1/a^d)$).
The suppression in the sign problem in the saturation regime was also seen in the work of \Ref{FDF}.

Another interesting fact is that $\<\cos\theta\>$ drops from $1$ at $\mu\approx m_\pi/2$ -- as chiral perturbation theory predicts (see, for example,  \Ref{LSV} and its references). Importantly, we point out that above $\mu/\sqrt{\lambda}\simeq 1.0$ the average sign is already very close to zero and so there is a severe sign problem there.

\begin{figure}[h!]
\centerline{
\includegraphics[width=15cm,height=10cm]{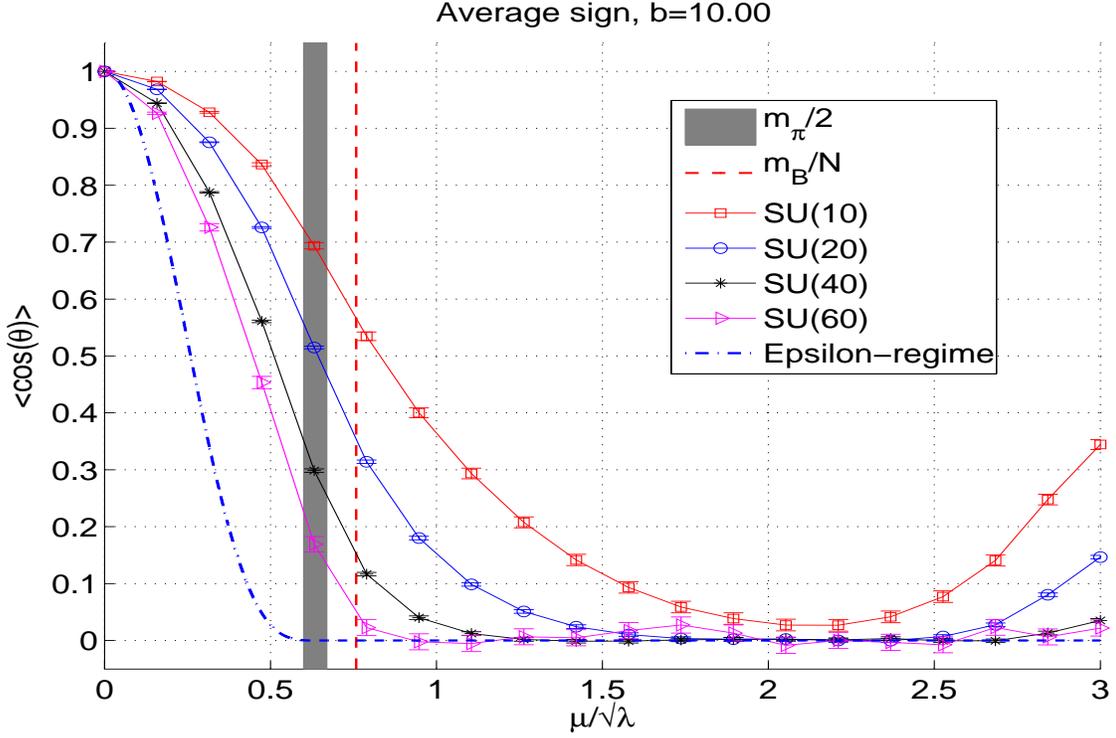}
}
\caption{Comparing the average sign of $SU(N)$ with $N=10,20,40$ and $60$ for $b=10.00$. In a dash-dot(blue) line is the average sign of the epsilon regime.}
\label{AvSign2}
\end{figure}

\begin{figure}[h!]
\centerline{
\includegraphics[width=12cm]{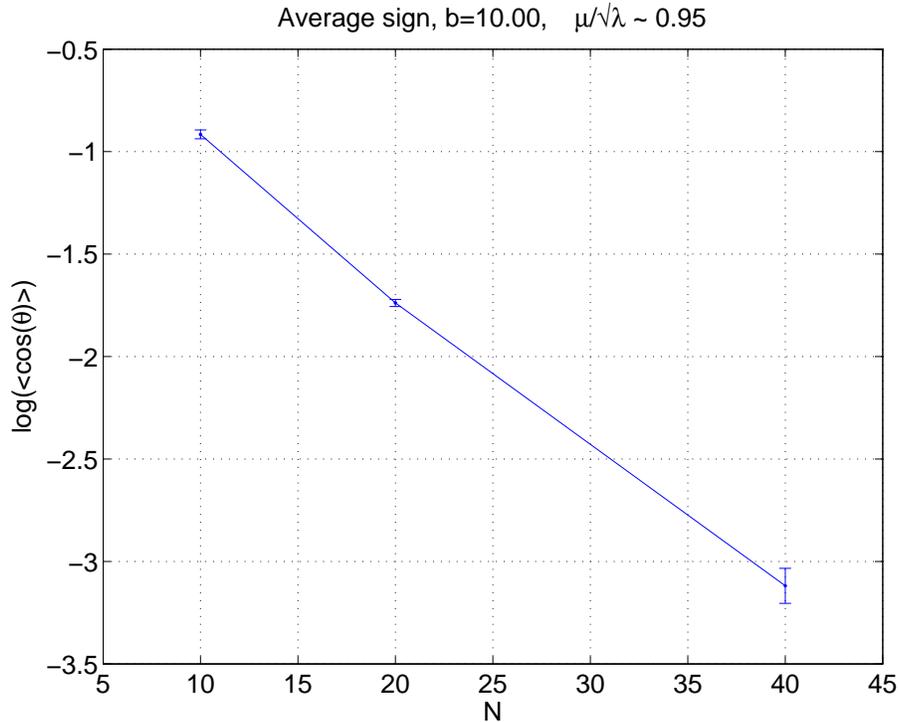}
}
\caption{A logarithmic plot of $\<\cos(\theta)\>$ versus $N$ at $b=10.00$ and $\mu/\sqrt{\lambda}\simeq 0.95$.}
\label{AvSign3}
\end{figure}

We now turn to study the way $\<\cos\theta\>$ changes with $N$. Since the large $N$ limit is the thermodynamic limit in the EK theory, we expect the sign problem to become worse as $N$ increases. This expectation is confirmed in Fig.~\ref{AvSign2} where we plot the average sign for $b=10.00$ and for the gauge groups $SU(10)$, $SU(20)$, $SU(40)$ and $SU(60)$.
We see that while there is no serious sign problem in $SU(10)$ for any value of $\mu$, for $SU(40)$ the average sign approaches zero at around $\mu/\sqrt{\lambda}\simeq 1.25$, and for $SU(60)$ the average sign is zero already at $\mu/\surd\lambda\simeq 0.75$. The way the average sign decreases with $N$ is expected to be exponential if $\mu>m_\pi/2$. This can be easily argued if a different definition of an average sign is used (for example, see \Ref{LSV}). Because we do not use the definition of \Ref{LSV}, we explicitly check how {\em our} definition for the sign changes with $N$. For that purpose we fix $\mu/\sqrt{\lambda}\simeq 0.95$ and measure $\<\cos\theta\>$ for $b=10.00$ and $N=10,20,40$. The results are presented on a logarithmic plot  in Fig.~\ref{AvSign3} and confirm that the sign drops approximately exponentially with $N$.

Before we proceed we wish to remark on the way the average sign behaves for $\mu<m_\pi/2$. Chiral perturbation theory gives different predictions for this behavior, depending on the value of the quark mass, the chemical potential and the volume. For example, in the thermodynamic limit, taken at fixed quark mass, the sign should freeze at one for all $\mu\le m_\pi/2$ \cite{SV}. In the so-called `epsilon regime', where the quark mass and $\mu$ are decreased when one takes the thermodynamic limit, the average sign approaches a smooth function of the variable $z\equiv \left(2\mu/m_\pi\right)^2$, that is equal to unity at $z=0$ and to zero at $z=1$. Therefore, interpreting the large-$N$ limit as the thermodynamical limit of our system, we expect that the large-$N$ limit of of $\<\cos\theta\>$ will be $O(1)$ for $z<1$ and zero for $z\ge 1$. From Figs.~(\ref{AvSign2})--(\ref{AvSign3}) it is fairly clear that for $z>1$ our results are consistent with this expectation.  The situation for $\mu<m_\pi/2$ is less clear since we do not know whether our quark mass and the values of $\mu$ that we study are inside the epsilon regime or not (for that we would need to measure $f_\pi$). Nonetheless, the fact that the sign clearly drops with $N$ suggests that we are either inside or close to the epsilon regime in that part of the phase diagram. Thus, as a guide to the eye we plot in Fig.~\ref{AvSign2} the average sign in the epsilon regime for our system. Here we assume that we are close enough to the continuum limit, and so we plot the quenched averaged sign $\<e^{4i\theta}\>_{\rm quenched}.$\footnote{I.e.~we set $p=2$ in Eq.~(48) of \Ref{LSV}} The factor of $4$ in the exponent  reflects the fact that we use naive fermions in our numerical studies;  in the continuum limit these fermions are four-fold doubled which means that our Dirac determinant $\det\left( D_{\rm lattice}\right)$ factorizes to $\left(\det D_{N_f=1}\right)^4$, where $D_{N_f=1}$ is the Dirac operator of a single flavor theory in the continuum limit.

\section{Testing $Z_N$ averaging.}
\label{ZN_test}

We now proceed to test the $Z_N$ averaging procedure by calculating the average sign of $\left(\det D\right)_Z$ for the same parameters presented in Fig.~\ref{AvSign2}. We denote this quantity by $\<\cos\theta\>_Z$, although we emphasize that what is measured is the sign of $\left(\det D \right)_Z$ and not the $Z_N$ average of $\cos \theta$. 

\begin{figure}[t]
\centerline{
\includegraphics[width=15cm,height=10cm]{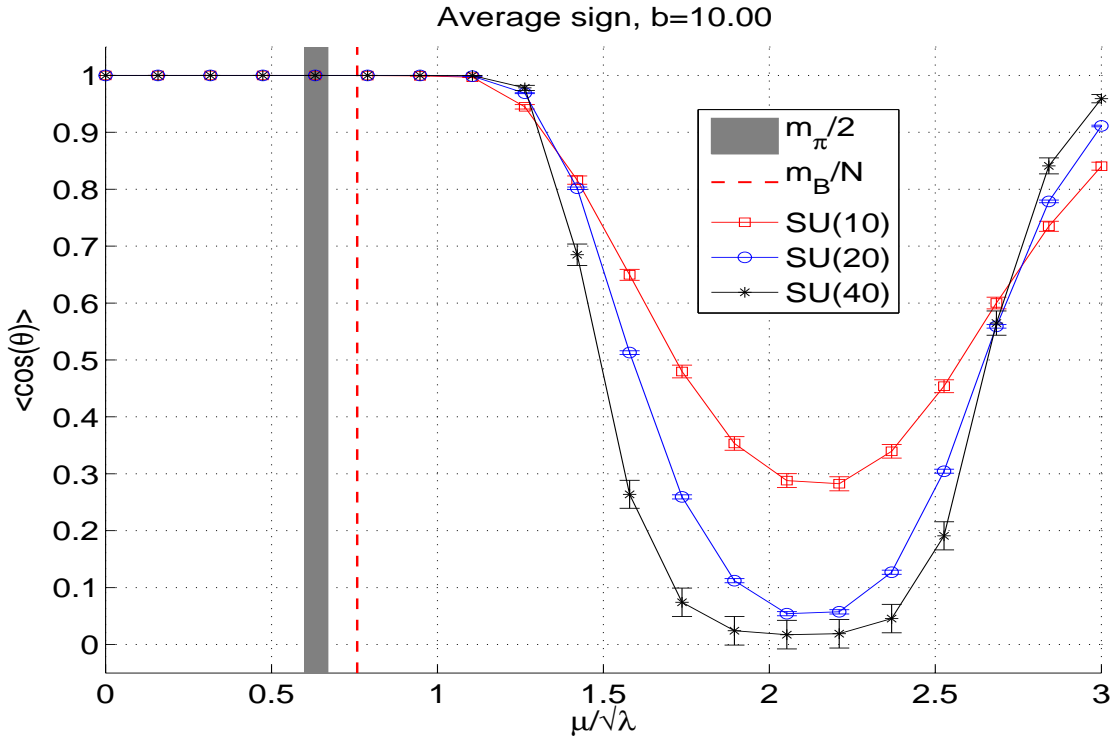}
}
\caption{The $Z_N$-averaged sign, $\left\<\cos(\theta)\right\>_Z$, for $b=10.00$ and $N=10,20,40$.}
\label{AvSign4}
\end{figure}

We present the results in Fig.~\ref{AvSign4} where we see that  $Z_N$-averaging makes a dramatic difference: throughout the range of chemical potentials $\mu/\sqrt{\lambda}\in [0, 1.2]$ the $Z_N$-averaged quantity $\left(\det D\right)_Z$ is real and positive. 
In fact, the $N$ dependence of $\<\cos\theta\>_Z$ suggests that this positivity will persist until about $\mu/\surd\lambda\simeq 1.25$, a regime that includes, as a subset, all of the hadronic phase. This is most clearly seen by comparing Fig.~\ref{AvSign4} with Fig.~\ref{AvSign2} --- there is a wide regime with $\mu>m_\pi/2$ where without $Z_N$ averaging the sign is exponentially small but  with $Z_N$ averaging it is equal to one. Thus we see that, in that range, $Z_N$ averaging solves the sign problem of our theory.

The absence of sign fluctuations below $\mu/\surd\lambda \simeq 1.25$ does not mean that there are no sign fluctuations for all $\mu$. In fact, for $SU(40)$ there is a severe sign problem beyond the hadronic phase, in the range $\mu/\sqrt{\lambda}\in[1.8,2.4]$.\footnote{Our statistical errors in Fig.~\ref{AvSign4} are larger compared to those presented in Fig.~\ref{AvSign2} since we were able to calculate $\<\cos\theta\>_Z$ for a fraction of our configurations in the $SU(40)$ case (recall that for each configuration we need to evaluate $\det D$ for each of the $40^2=1600$ terms in the sum in \Eq{ZN_def}). Thus, our resources allowed us to only do so for $5000$ of the $105000$ configurations.} This is a real  sign problem (i.e.~it is not `just' a silver-blaze problem) because it appears when observables start to be $\mu$-dependent. Also note that {\em this} sign problem precedes the lattice saturation of the EK theory, where the sign fluctuations goes away again.

\begin{figure}[h!]
\centerline{
\includegraphics[width=15cm,height=8cm]{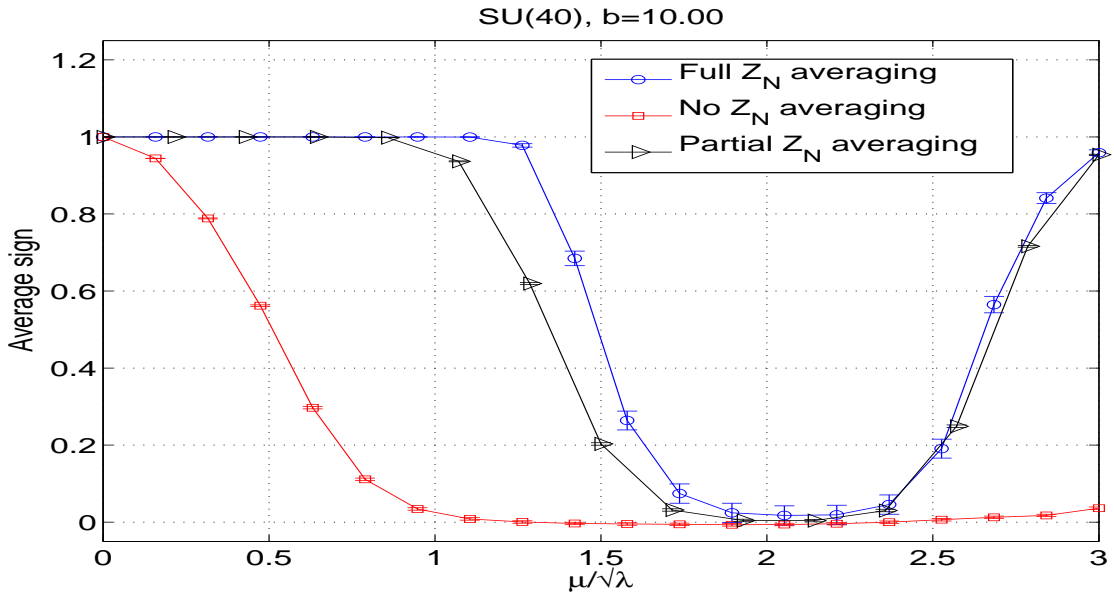}
}
\caption{The average sign with different versions of averaging. Here $b=10.00$ and the gauge group is $SU(40)$.}
\label{AvSign5}
\end{figure}

What happens if we only perform a partial $Z_N$ averaging (see discussion at the end of \Sec{ZN})? We present a comparison of the two type of averaging for $N=40$ and $b=10.00$ in Fig.~\ref{AvSign5}.
\begin{figure}[h!]
\centerline{
\includegraphics[width=15cm,height=10cm]{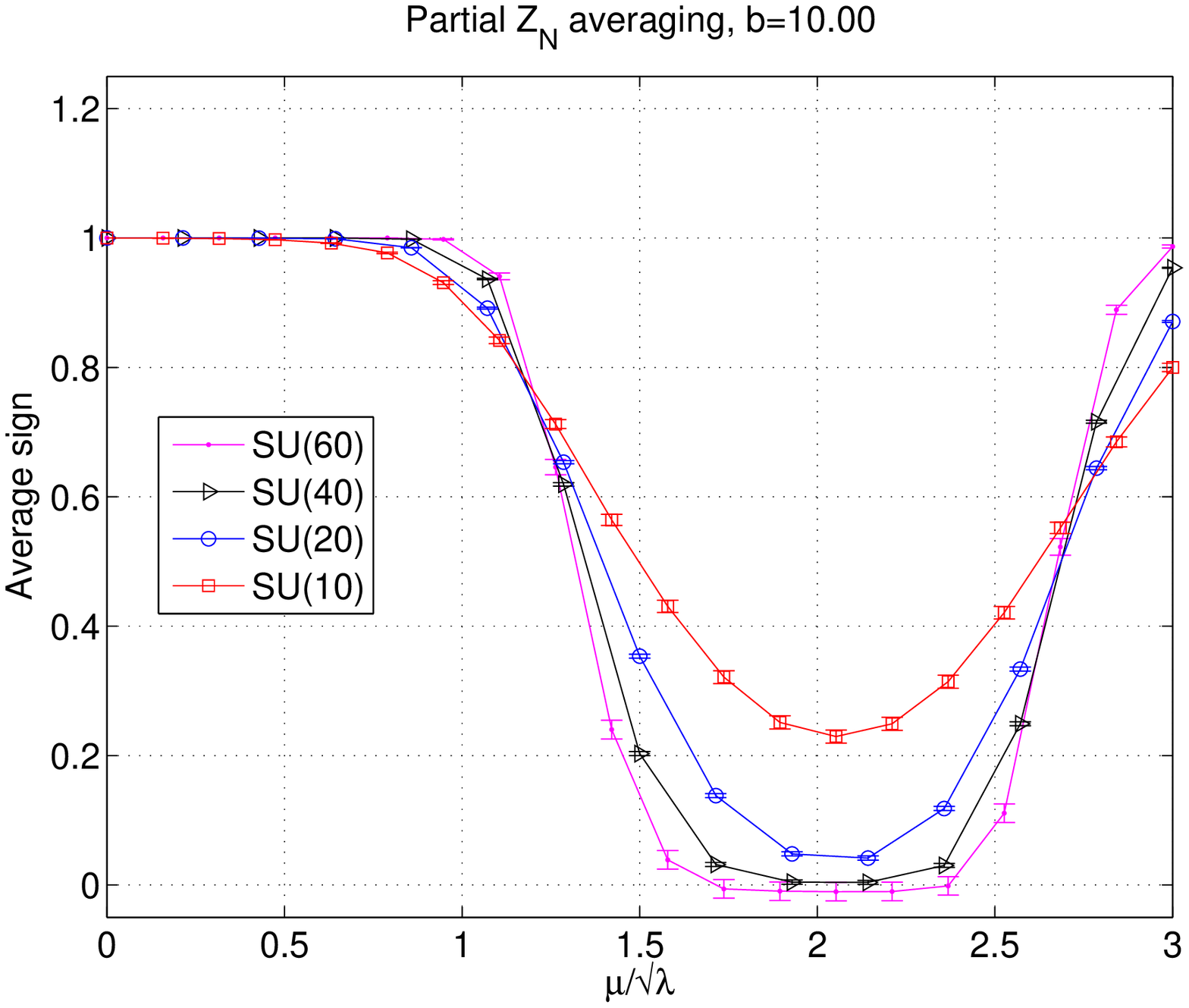}
}
\caption{The average sign with partial $Z_N$ averaging. Here $b=10.00$ and the gauge groups are $SU(10)$, $SU(20)$, $SU(40)$, and $SU(60)$.}
\label{AvSign6}
\end{figure}
The fact that the two averaged signs are nearly the same (contrast their behavior with the non-$Z_N$-averaged sign in the figure) means that a partial $Z_N$ averaging works almost as well as the full one, and indeed we shall use it when our resources cannot accomplish the latter. An interesting point appears when we compare the results of partial $Z_N$ averaging for various values of $N$ in Fig.~\ref{AvSign6}. There, we see that, similarly to the full $Z_N$ averaging case, the sign {\em grows} towards $1$ for $\mu/\surd\lambda\stackrel{<}{_\sim} 1.25$ and $\mu/\surd\lambda \stackrel{>}{_\sim} 2.75$. This means that in these ranges of $\mu$, partial $Z_N$ averaging removes the sign fluctuations from the averaging of $\det D$ in a way which is as good as the one provided by full $Z_N$ averaging. 

\bigskip

Let us now ask whether $Z_N$ averaging actually improves the signal to noise ratio of physical observables. Since unquenched observables always involve the calculation of $\<\det D\>$, we ask what is the improvement that $Z_N$ averaging has to offer in the calculation of the latter (or more precisely  in  the calculation of the free energy $F\equiv \frac1{N} \, \log \<\det D\>$). 

\begin{figure}[h!]
\centerline{
\includegraphics[width=15cm]{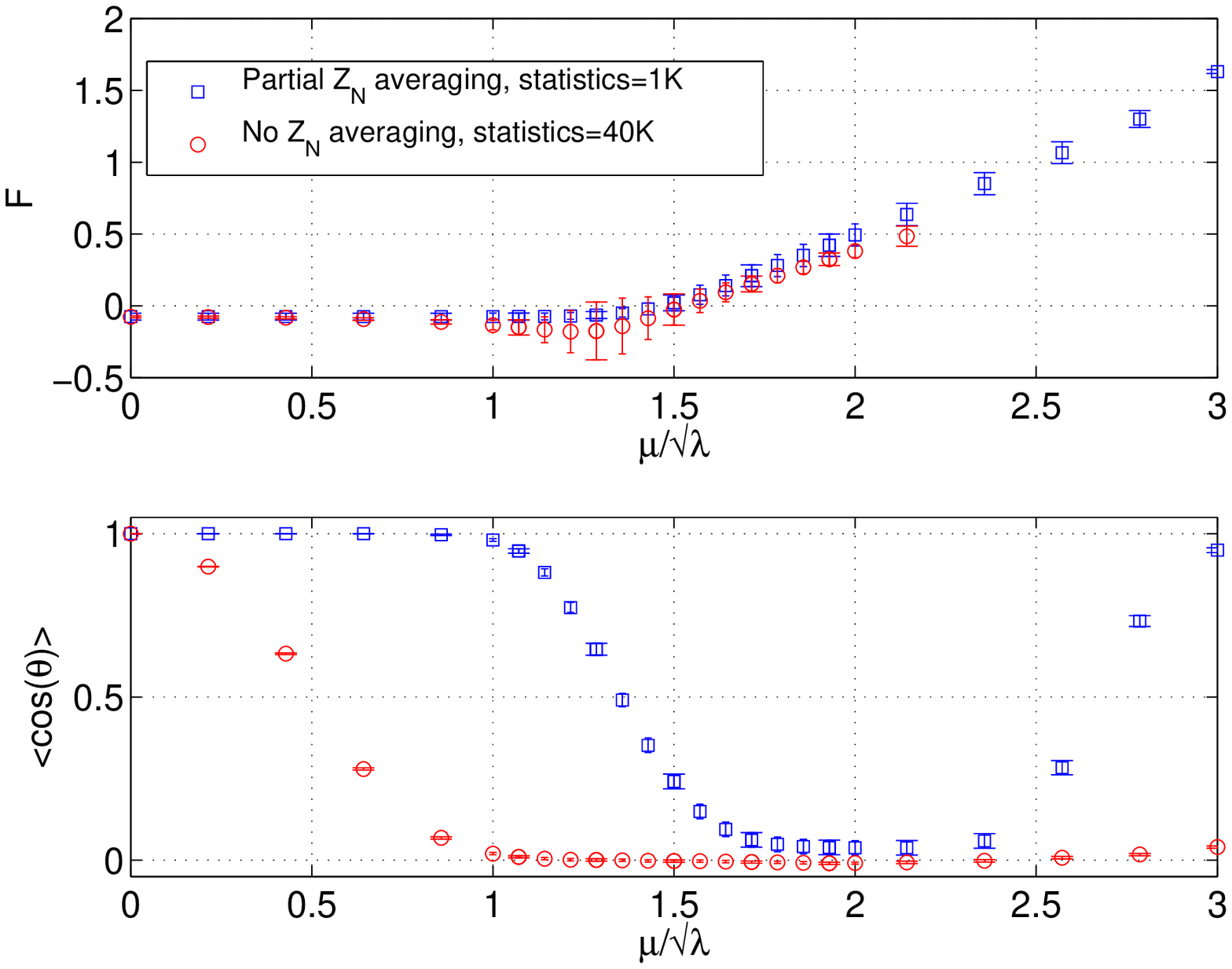}
}
\caption{A test of the efficiency of $Z_N$ averaging for $SU(40)$ and $b=10.0$. Squares(blue) show the result of method no.~(2) while circles(red) those of method no.~(1). In the upper panel we show $F$ and in the lower one the average sign.}
\label{ZN_efficiency_F}
\end{figure}

\begin{figure}[h!]
\centerline{
\includegraphics[width=15cm,height=8cm]{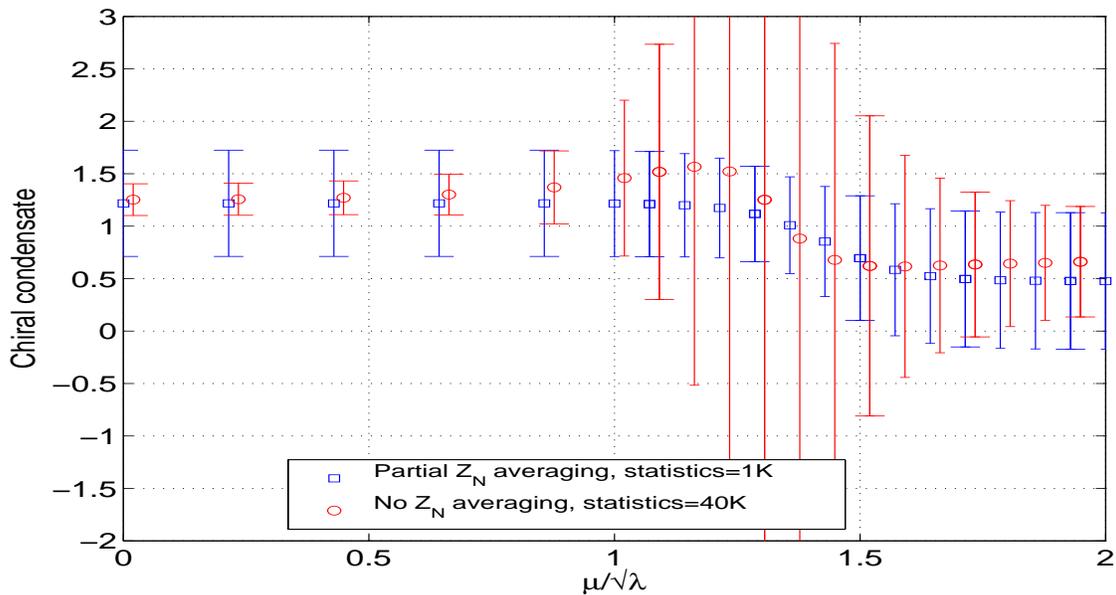}
}
\caption{A test of the efficiency of $Z_N$ averaging for $SU(40)$ and $b=10.00$ for the chiral condensate. Squares(blue) show the result of method no.~(2) while circles(red) those of method no.~(1). The data obtained with method no.~(1) is slightly shifted on the $x$-axis, to distinguish it from the data of method no.~(1).}
\label{ZN_efficiency_pbp}
\end{figure}

To answer this question we check which of the following methods gives the smallest statistical error on $F$, provided we fix the computational effort. 
\begin{enumerate}[(1)]
\item No $Z_N$ averaging. Here we average a set of $M$ gauge configurations.
\item With partial $Z_N$ averaging. Here we use a set of $M/N$ gauge configurations. Recall that for given gauge configuration, we calculate the determinant $N$ times (see \Sec{ZN}).
\end{enumerate}
We perform our check for $N=40$, $M=40000$, and $b=10.0$ -- a case where the sign problem can be relatively severe. 
The results of this check are given in Fig.~\ref{ZN_efficiency_F}.
 In the upper panel we show $F$, as obtained with both methods, and for convenience we show the corresponding average signs in the lower panel of the figure. A quick look at the figure shows that $Z_N$ averaging can be useful. In fact, note that we do not show the results for $F$ from method no.~(1) when $\mu/\surd\lambda>2.25$. This is because the severe sign fluctuations that appear in that method make the average of $\det D$ negative. In contrast, the average sign of method no.~(2) is away from zero in the same regime and so $\left(\det D\right)_{{\rm partial}-Z}$ is positive and real there. 
The most important point, however, is that for $\mu/\surd\lambda \le 1.25$ and $\mu/\surd\lambda\ge 2.75$, the sign is exponentially small in $N$ with method no.~(1) while it is increasing towards unity with method no.~(2). Since the numerical cost of method no.~(2) is linear in $N$, we see that it provides a gain which is exponential in $N$.\footnote{The fact that in method no.~(2) one has less independent configurations might mean that its statistical error can be larger, but at sufficiently large-$N$ and a moderately large number of configurations, the sign problem will always make method no.~(2) preferable. This is already seen for $SU(40)$ in Fig.~\ref{ZN_efficiency_F} in the range between $\mu/\surd\lambda\in[1,1.25]$.} 
To show the effect of the errors in the averages of $\det D$ on unquenched observables like $\<\bar\psi\psi\>$, we plot the results obtained for $\<\bar\psi\psi\>$ with the two methods in Fig.~\ref{ZN_efficiency_pbp}. Clearly $Z_N$ averaging is useful here. 

In the next section we show more results for $F$ and $\<\bar\psi\psi\>$.
 These were obtained with $Z_N$ averaging and for much larger statistical
 samples than the ones we discuss above.

\section{The free energy and the chiral condensate}
\label{physical}

\begin{figure}[bt]
\centerline{
\includegraphics[width=17cm]{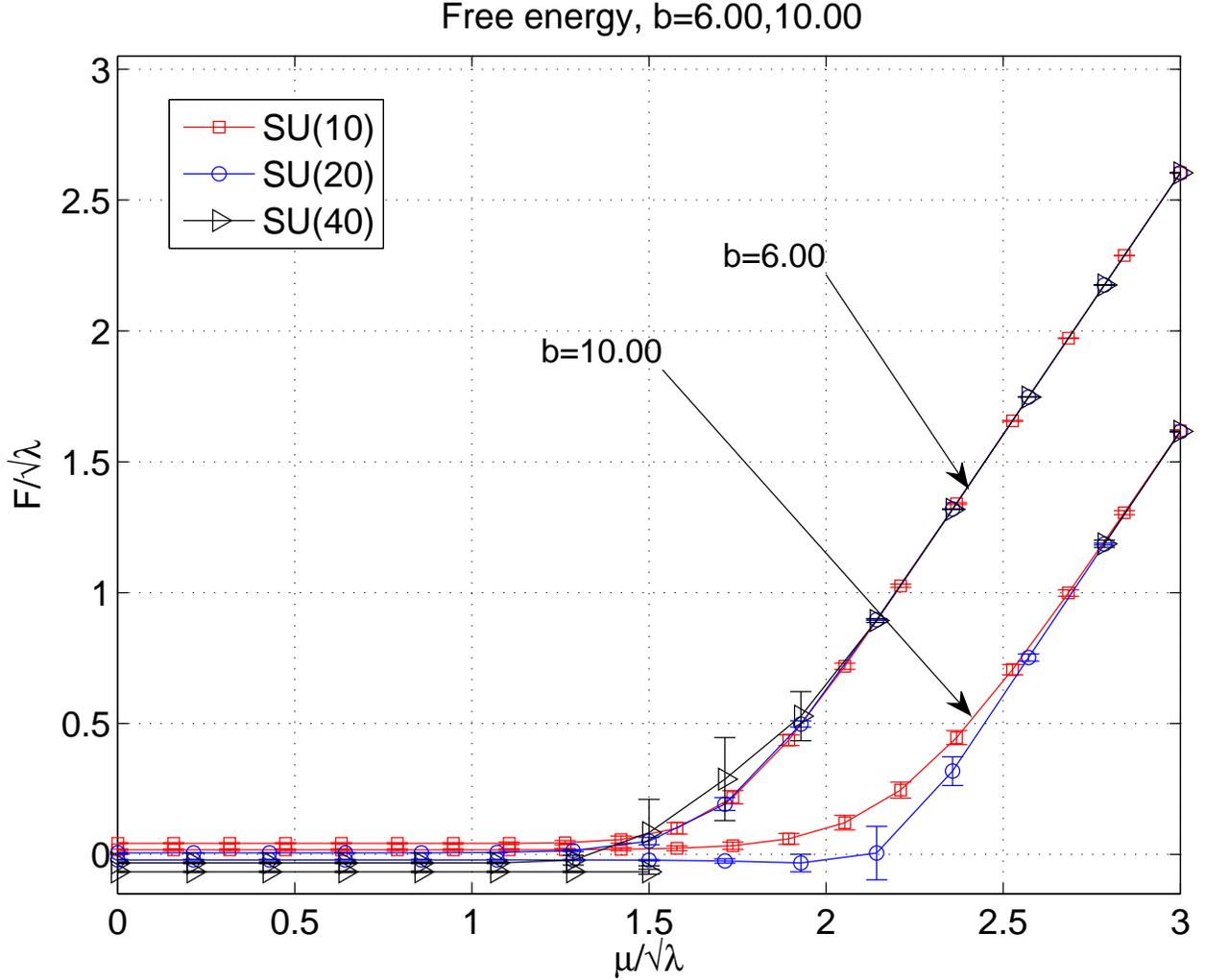}
}
\caption{The $Z_N$-averaged free energy density $F_Z$ (see \Eq{F1}) of $SU(10),SU(20)$, and $SU(40)$ at $b=6.0$ and $10.0$. The calculation for the $SU(40)$ case was done with partial $Z_N$-averaging. To guide the eye we connect the data with solid lines.}
\label{FB}
\end{figure}

In this section we present several results of physical interest, beginning with the way the free energy $F$ behaves as a function of $\mu$. We calculated $F$ using $Z_N$ averaging and present the results for $b=6.0,10.0$ and $N=10,20$  in Fig.~\ref{FB}. Performing $Z_N$ averaging for $N=40$ was too challenging for our resources and so in that case we only used partial $Z_N$ averaging. Also, while the $SU(40)$ results are similar to those of $SU(10)$ and $SU(20)$ for $\mu/\sqrt{\lambda}<1.5$, the errors in the regime of  the `true' sign problem  make the data not useful there -- this is  why we do not present the data for $b=10.00$ in the regime $\mu/\surd\lambda\in [1.5,2.75]$. Comparing this figure to the upper panel of Fig.~\ref{ZN_efficiency_F}, we see that, beyond the hadronic phase, when the true sign problem is at its peak, even $Z_N$ averaging does not help. This is reflected by the fact that while the smaller statistical sample used to produce Fig.~\ref{ZN_efficiency_F} gave positive values of $\<\det D\>$,  the statistical sample we use in this section, which is $100$ times larger, resulted in negative values of $\<\det D\>$ in the same regime. 

There are two additional  important points to take away from Fig.~\ref{FB}. First, we see that the free energy is completely independent of $\mu$ throughout the hadronic phase, as we expect physically for $T=0$. This is another way of seeing that the silver-blaze problem is absent from our calculation. Second, we see that $F$  becomes $\simeq 2\mu$ for large value of $\mu$. This is the lattice saturation we discussed above (see \Eq{sat}). 

\bigskip

\begin{figure}[h!]
\centerline{
\includegraphics[width=17cm,height=15cm]{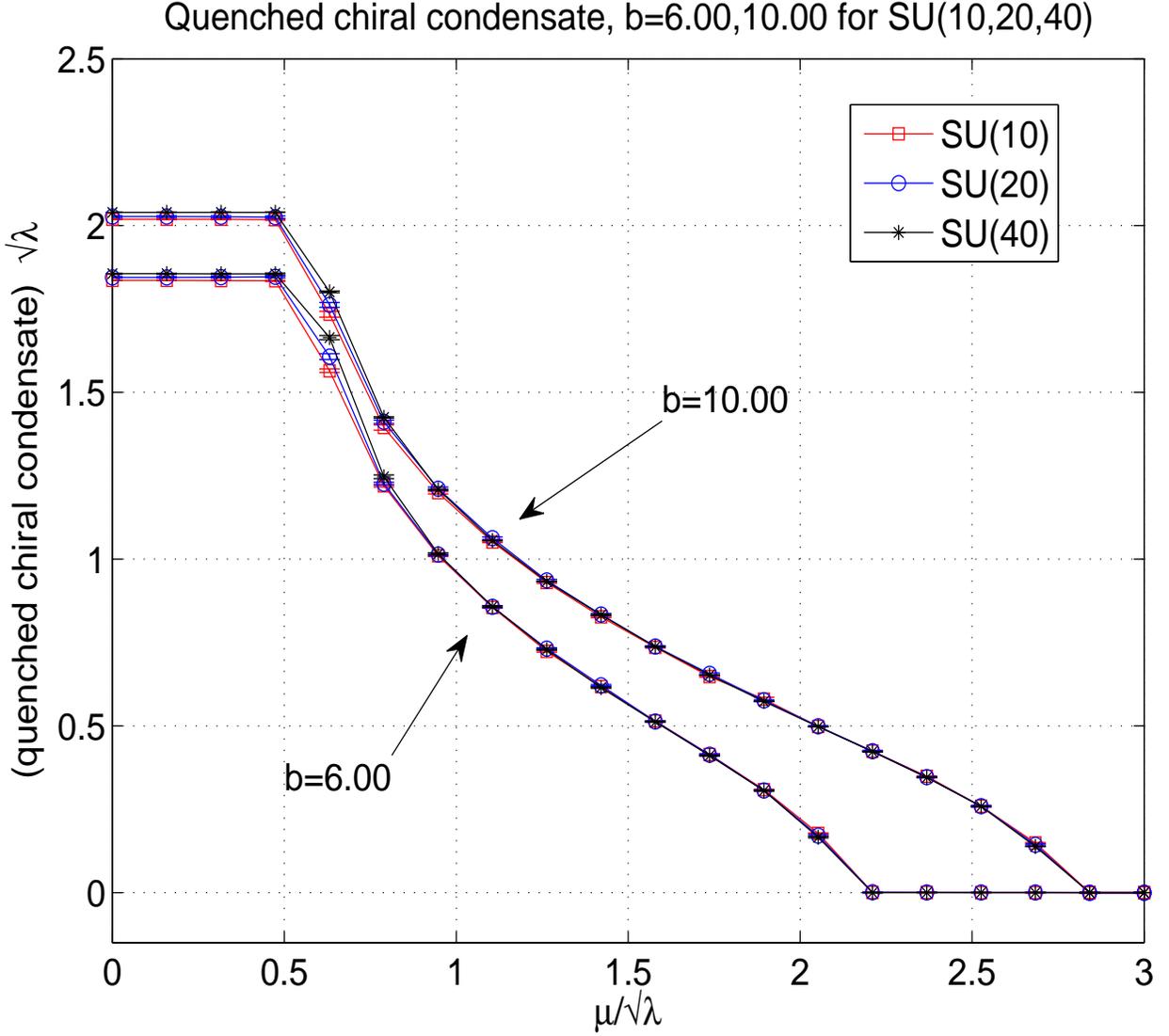}
}
\caption{The quenched quark condensate for $N=10,20,40$ and $b=6.00,10.00$.
}
\label{SQ}
\end{figure}

To emphasize that the results in the preceding sections are truly unquenched  we present the behavior of the quenched condensate $\<\bar\psi\psi\>_{\rm quenched}$  in Fig.~\ref{SQ}. 
The sharp change in $\<\bar\psi\psi\>_{\rm quenched}$ at around $m_\pi/2$ is the same as seen in other quenched lattice calculations, and it is a nonsensical result reflecting the  mutilation that the quenched prescription causes to the gauge theory at nonzero $\mu$. Comparing Fig.~\ref{SQ} to the plot of the free energy, Fig.~\ref{FB}, we see that the drop in the former happens at values of $\mu$ where the free energy is still independent of $\mu$. 

\begin{figure}[h!]
\centerline{
\includegraphics[width=15cm,height=10cm]{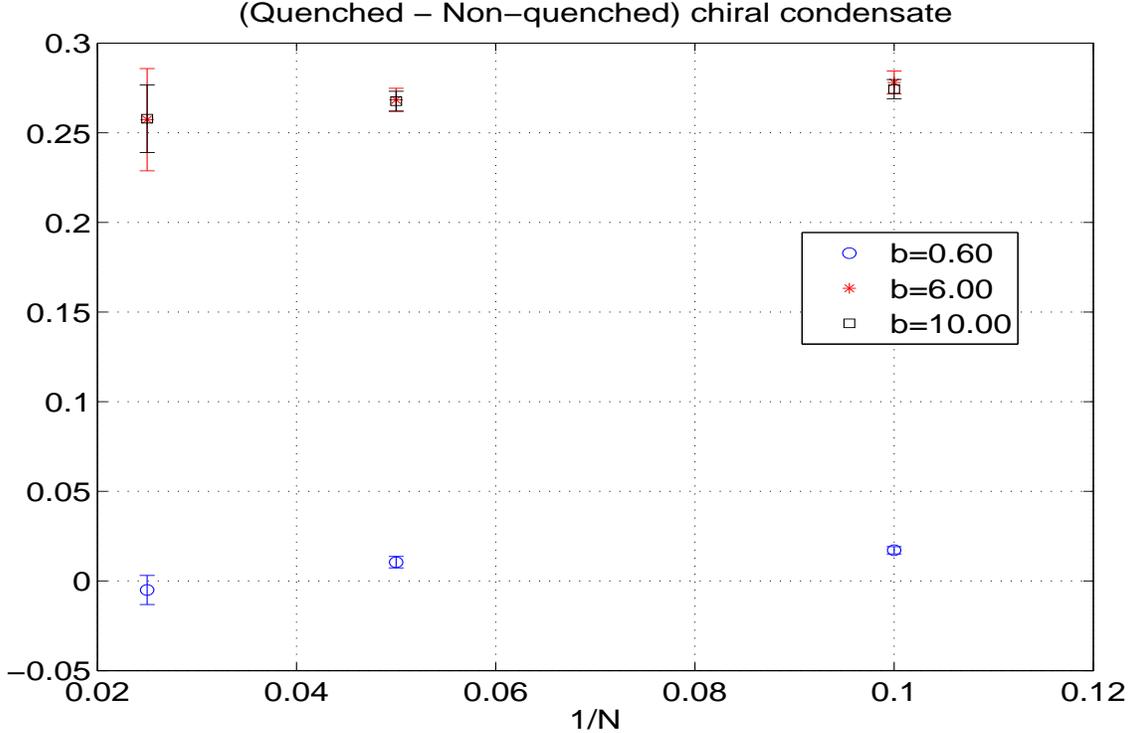}
}
\caption{The difference between the quenched and unquenched chiral condensate for $b=0.60,6.00,10.00$ versus $1/N$ for $\mu=0$.
}
\label{Q_vs_NQ}
\end{figure}

We attempted to use $Z_N$ averaging and calculate the {\em unquenched} condensate, but for $SU(40)$ this proved too costly for our resources.\footnote{The bottle neck was not the generation of configurations, but rather the repeated calculation of the observables for each gauge configuration. For example, analyzing $10^4$ measurements of $SU(40)$ for a single value of $\mu$ with $Z_N$ averaging would take around 555 hours on a $2.66$GHz CPU.} Instead, we calculated it using partial $Z_N$ averaging (see \Sec{ZN}). 
Let us now compare the quenched and unquenched condensates. 
We begin at $\mu=0$, where we expect both condensates to be equal when $N=\infty$. We test this in Fig.~\ref{Q_vs_NQ} where we plot the difference $\left(\<\bar\psi\psi\>_{\rm quenched}-\<\bar\psi\psi\>_{\rm unquenched}\right)/N$ for the three lattice couplings $0.60,6.0,10.0$ and versus $1/N$.
As the figure shows, while the difference indeed goes to zero for $b=0.60$, it does not seems to do so for $b=6.0$ and $10.0$. We interpret this as resulting from  working at too small values of $N$. Similar differences between quenched and unquenched averages at $\mu=0$ were also reported for the overlap operator in this system \cite{Rajamani}; it would be useful to understand this slow convergence to $N=\infty$, but this is beyond the goals of our current study.

We now turn to compare  the quenched and unquenched condensates at $\mu\neq0$. In this case, our discussion in \Ref{BY}  shows that the quenched approximation fails and that we should expect large $O(1)$ deviations (that do not go to zero at large $N$). While we explain this failure in \Ref{BY} let us think about it from the point of view of the numerical calculation. If  the quenched approximation was exact at large-$N$ it would mean that the operators $\bar\psi\psi$ and $\det D$ are classical operators that obey large-$N$ factorization since then we would have
\begin{equation}
\<\bar\psi \psi \times \det D\> \stackrel{N\to \infty}{=}  \<\bar\psi \psi\> \times \<\det D\>.\label{fac}
\end{equation} 
At nonzero $\mu$ we know that \Eq{fac} fails because quenching fails --- see the fictitious phase transition at $\mu=m_\pi/2$ observed in Fig.~(\ref{SQ}). Thus we expect correlations generated by the strong fluctuations that both $\bar\psi\psi$ and $\det D$ have within the complex plane. Put differently, the angles $\theta$ and $\alpha$ defined through Eqs.~(\ref{theta}--\ref{alpha})
now fully spread over the range $[-\pi,\pi)$ and can become correlated even when $N=\infty$. We will investigate this issue numerically in the next section. 

\begin{figure}[bt]
\centerline{
\includegraphics[width=17cm]{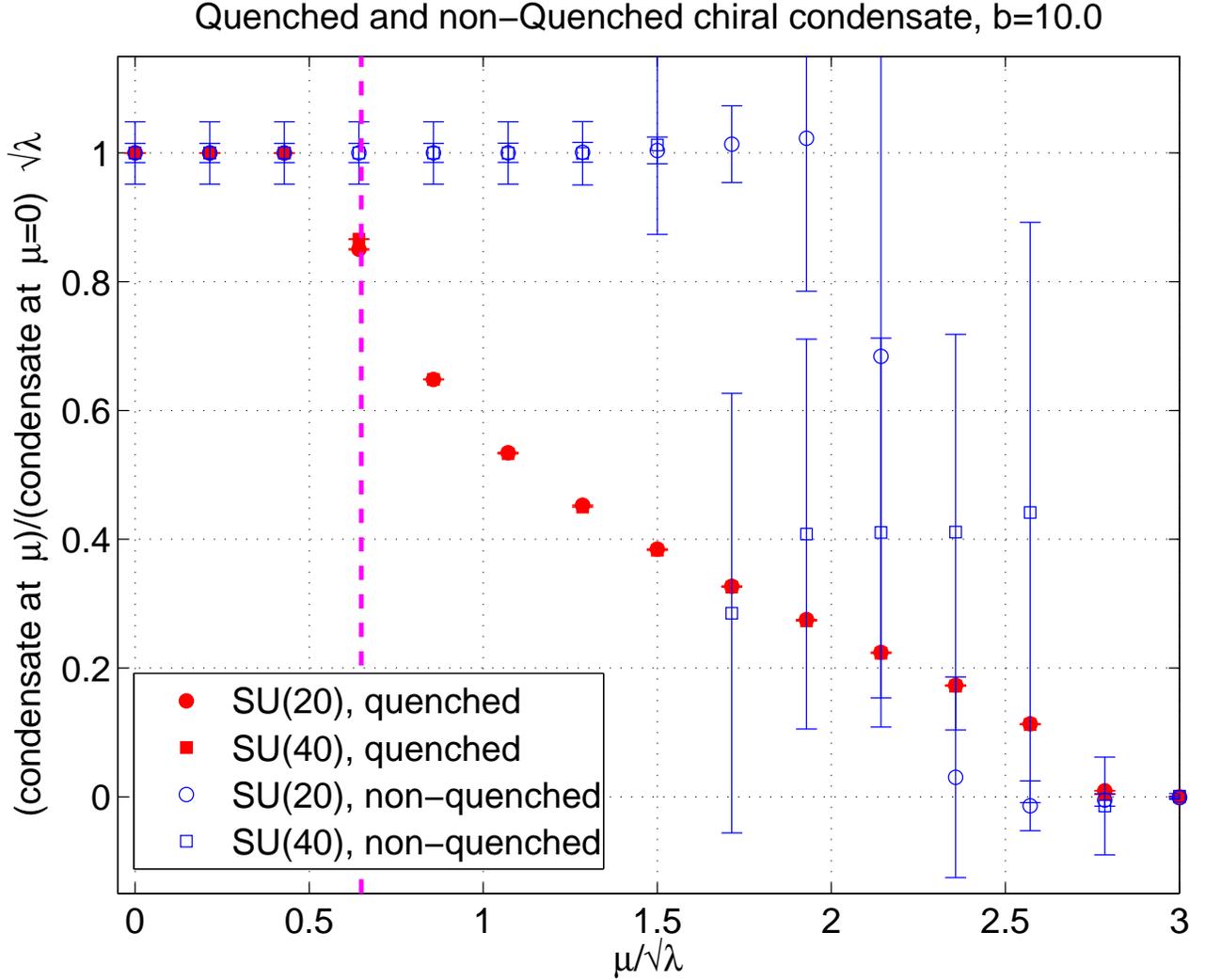}
}
\caption{The quenched and unquenched chiral condensate, normalized by their 
$\mu=0$ value, versus $\mu$. Here $b=10.00$ and we present results for $N=20,40$. The vertical magenta dashed line denotes our estimate for half the pion mass (see Table~\ref{lat_parm}).}
\label{Q_vs_NQ_mu}
\end{figure}

A direct measure of such correlations is the way the difference between the quenched and unquenched chiral condensates behave as a function of $\mu$. We already saw that at the values of $N$ that we work with, the $\mu=0$ condensates are different. This, however, we associated to significant $1/N$ corrections and distinguishing these from the $O(1)$ differences that we expect at nonzero $\mu$, is hard to do unambiguously. Nevertheless, we attempt to do so by presenting in Fig.~\ref{Q_vs_NQ_mu} the way the quenched and unquenched condensates, normalized by their $\mu=0$ value, change with $\mu$, for $b=10.0$ and $N=20$ and $40$.


There are a few useful observations we can make on Fig.~\ref{Q_vs_NQ_mu}. First, we see that the behaviors of the quenched and unquenched condensates as a function of $\mu$ strongly differ for chemical potentials that are above half the pion mass (denoted by the vertical magenta dashed line), as expected from chiral perturbation theory. Specifically, we see that despite the sharp change in $\<\bar\psi\psi\>_{\rm quenched}$ at $\mu\simeq m_\pi/2$, the physical unquenched condensate is unchanged throughout the hadronic phase and beyond. Second, we see that, within our statistical errors, the results for $N=20$ and $40$ are nearly on top of each other. Thus, combined with the theoretical expectations of \Ref{BY}, it does not seem likely that the differences between the quenched and unquenched averages that we see for $\mu\stackrel{>}{_\sim} m_\pi/2$ are $1/N$ effects. Finally, we see that the errors on the unquenched condensate greatly increase in the regime of the real `non-silver-blaze' sign problem (see Fig.~\ref{AvSign3}). 
All these facts are in agreement with physical expectations.


\section{Distributions in the complex plane}
\label{phasedetD}

This section has three goals. First we wish to examine the correlations between $\bar\psi\psi$ and $\det D$ that cause the failure of the quenched approximation. Second, we wish to see if our results for the distribution of $\theta$ (as defined by \Eq{theta}) varies with $\mu$ as predicted by chiral perturbation theory \cite{LSV}. In that reference, the authors showed that for $\mu<m_\pi/2$, $\theta$ is distributed as a periodic Gaussian, while for $\mu>m_\pi/2$ the distribution becomes a periodic Lorentzian. Also, the widths of these distributions go to infinity in the thermodynamic limit (which in our case is the large-$N$ limit). Third, we wish to see if the distribution of the eigenvalues of $D$ in the complex plane is consistent with chiral perturbation theory.

\begin{figure}[h!]
\centerline{
\includegraphics[width=10cm,height=7cm]{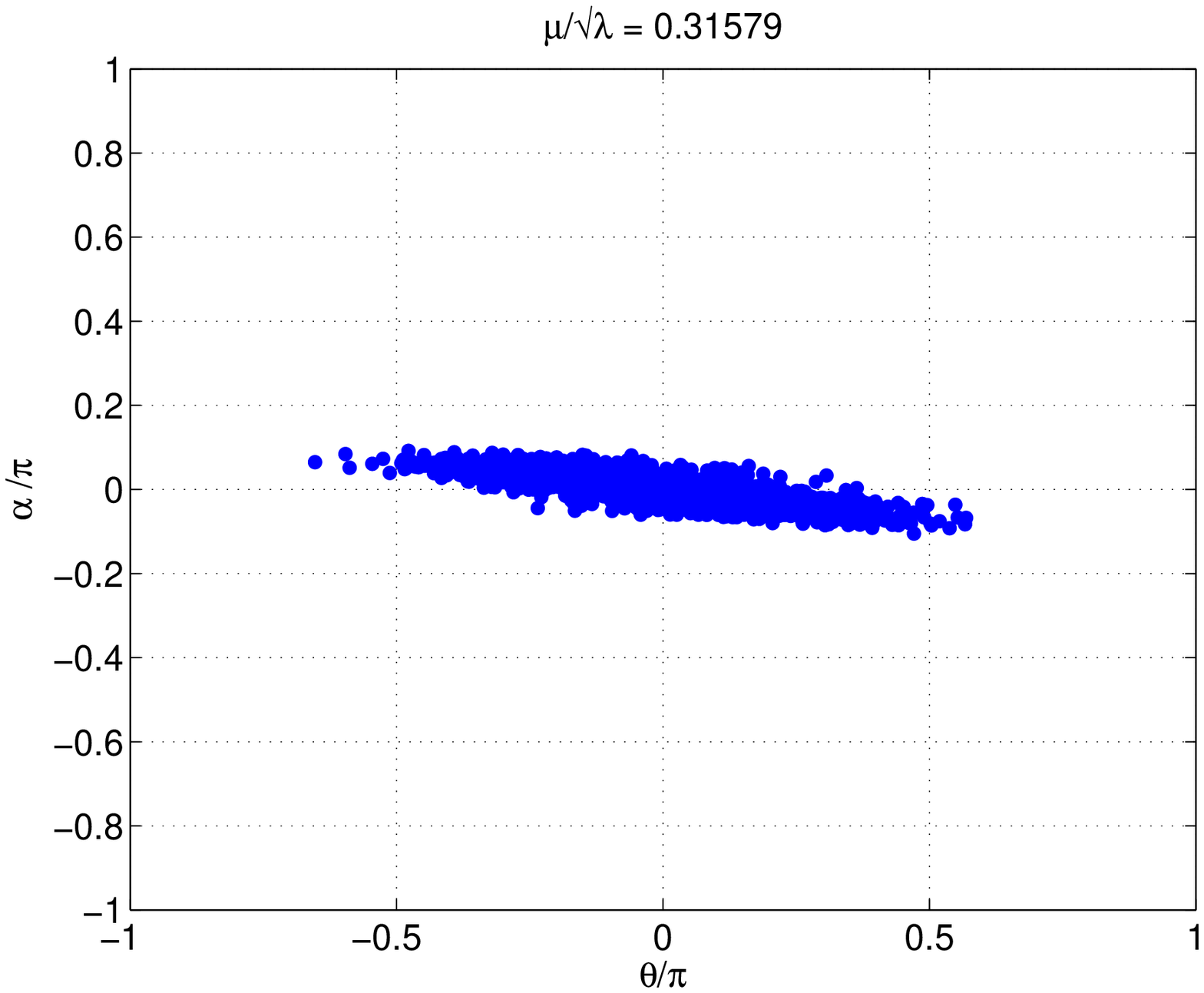}
\includegraphics[width=10cm,height=7cm]{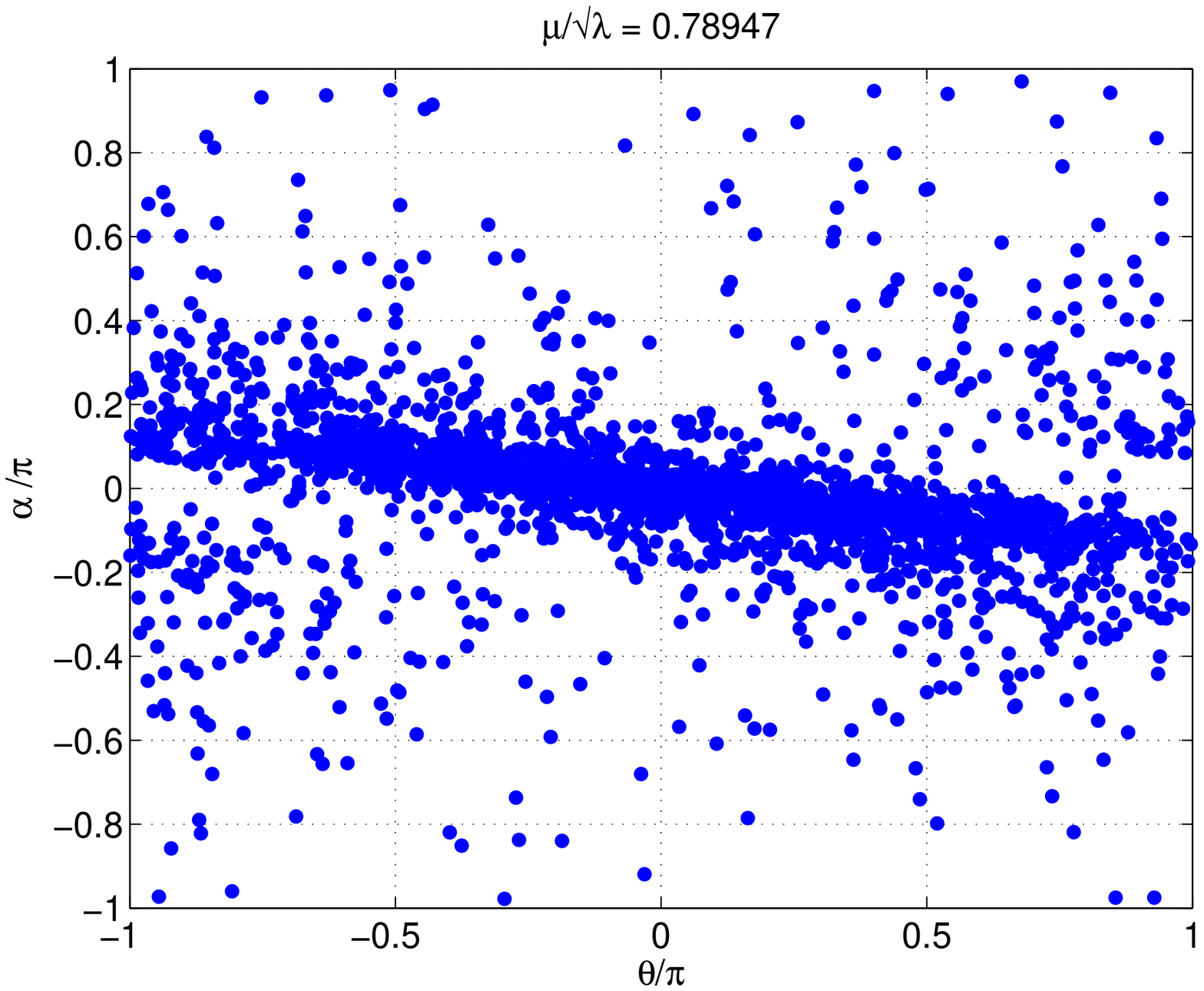}
}
\centerline{
\includegraphics[width=10cm,height=7cm]{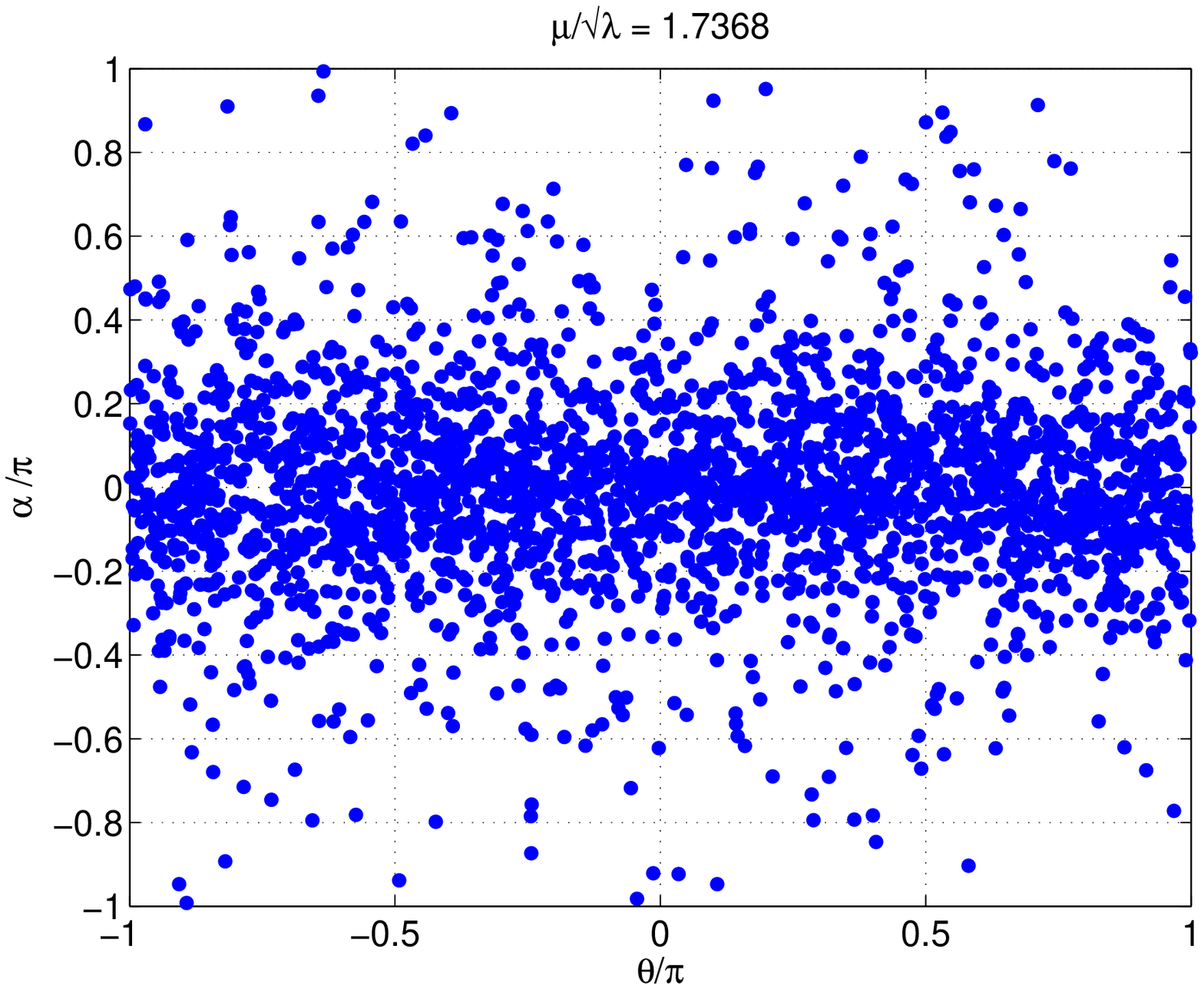}
\includegraphics[width=10cm,height=7cm]{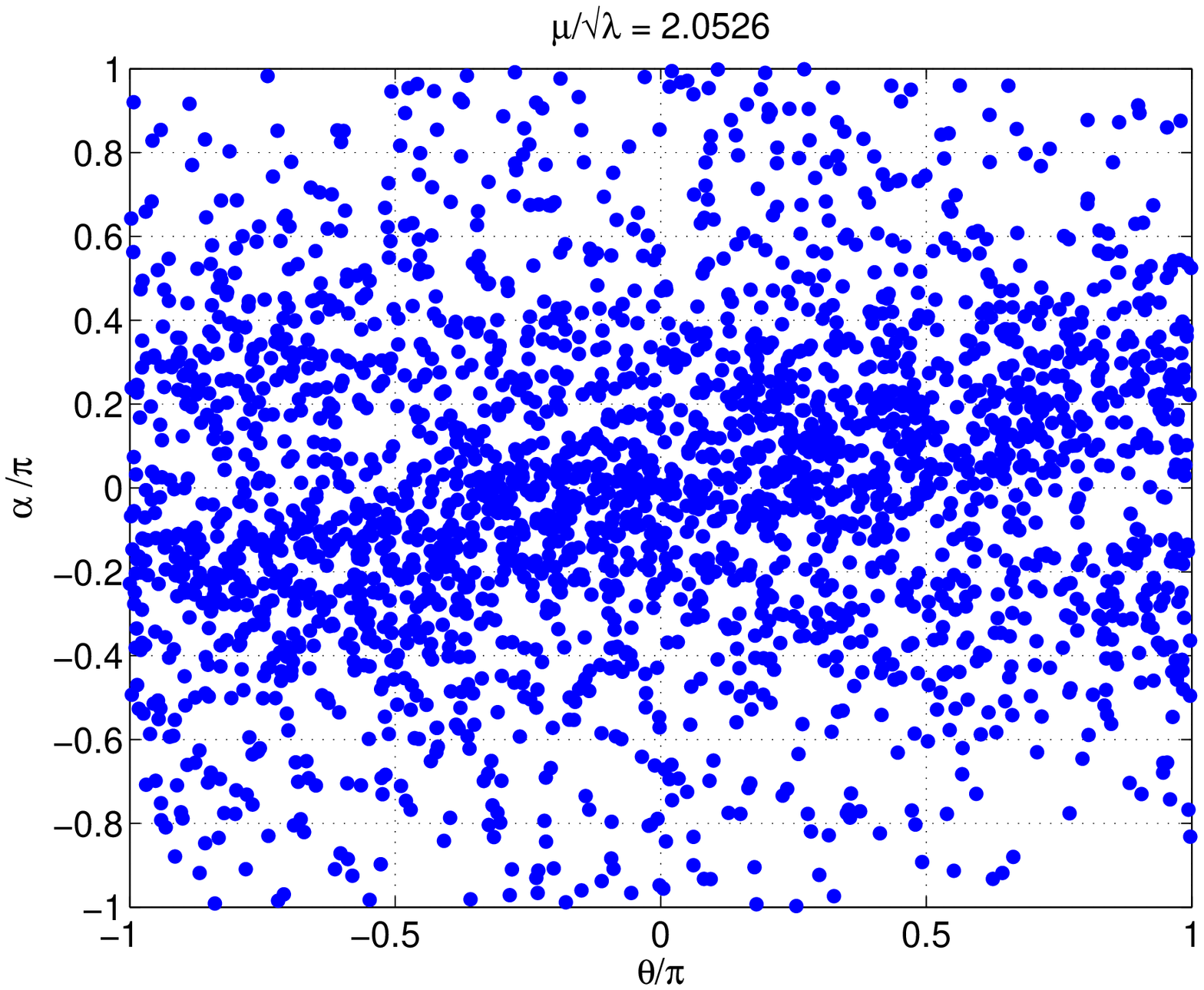}
}
\centerline{
\includegraphics[width=10cm,height=7cm]{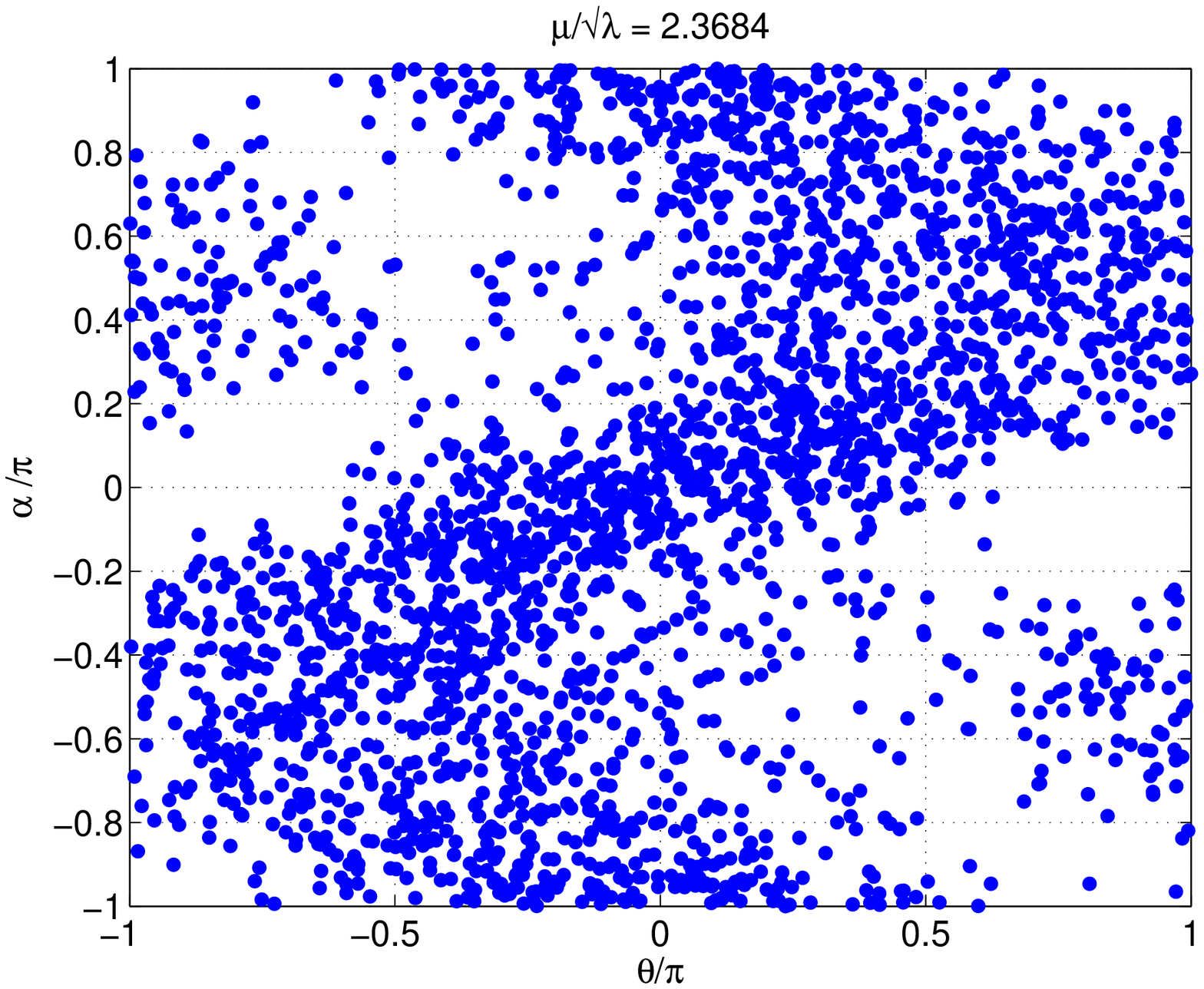}
\includegraphics[width=10cm,height=7cm]{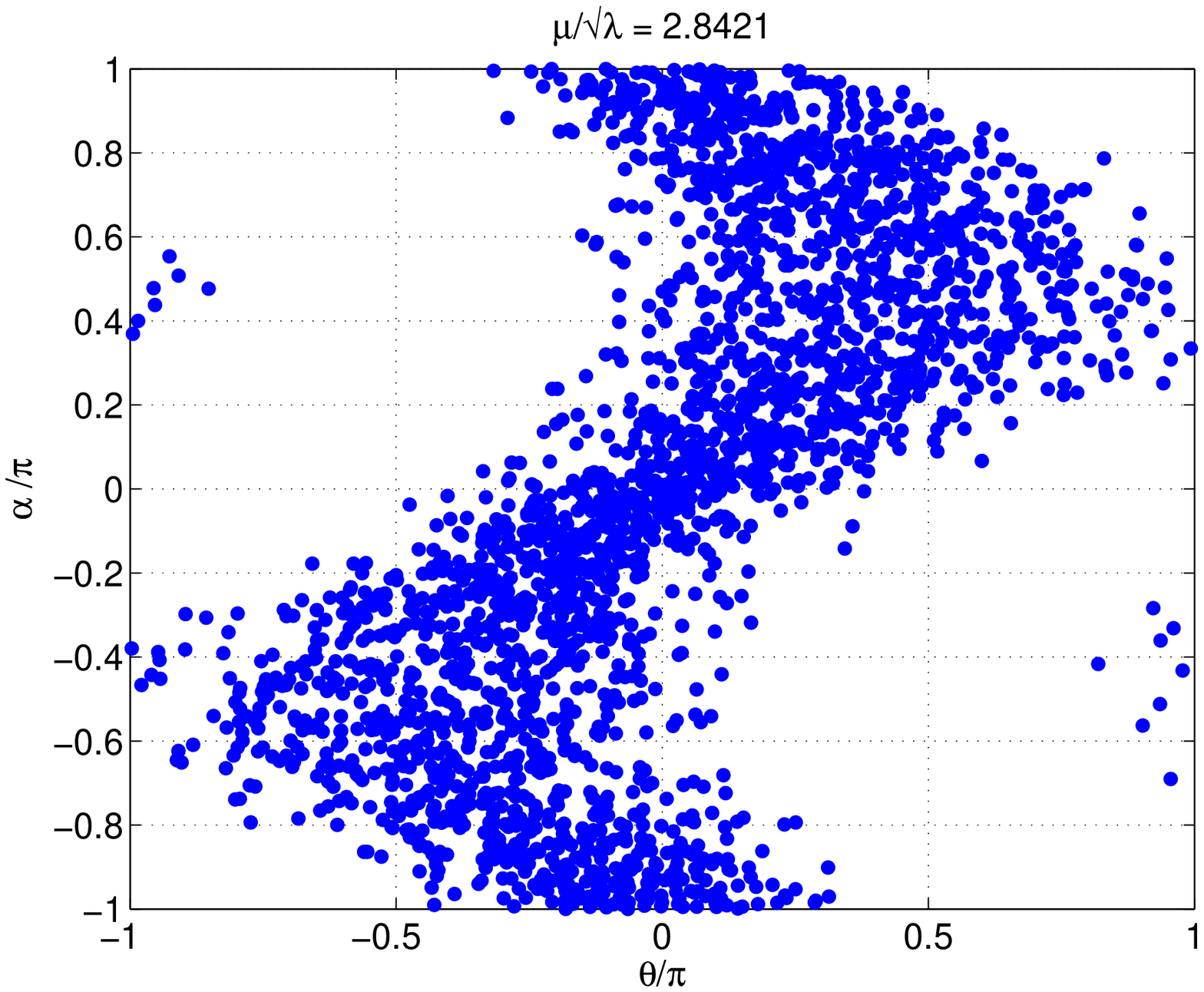}
}
\caption{The correlations between $\alpha$ and $\theta$ (see text) for $SU(20)$, $b=6.0$, and various values of $\mu$.
}
\label{corr1}
\end{figure}

We begin with analyzing the correlations between the phase $\alpha$ of $\bar\psi\psi$ and $\theta$, and present these for $SU(20)$, $b=6.0$, and different values of $\mu$ in Fig.~\ref{corr1}.
What we learn from the figure is that for most values of $\mu$ there is some correlation between $\alpha$ and $\theta$. Especially, it seems that the fluctuations of $\alpha$ and $\theta$ are not small, and that for $\mu>m_\pi/2$, they spread in the range $[-\pi,\pi)$. We note that this, however, does not mean that they are strongly correlated since their joint probability distribution can still be approximately separable.

\begin{figure}[h!]
\centerline{
\includegraphics[width=10cm,height=9.25cm]{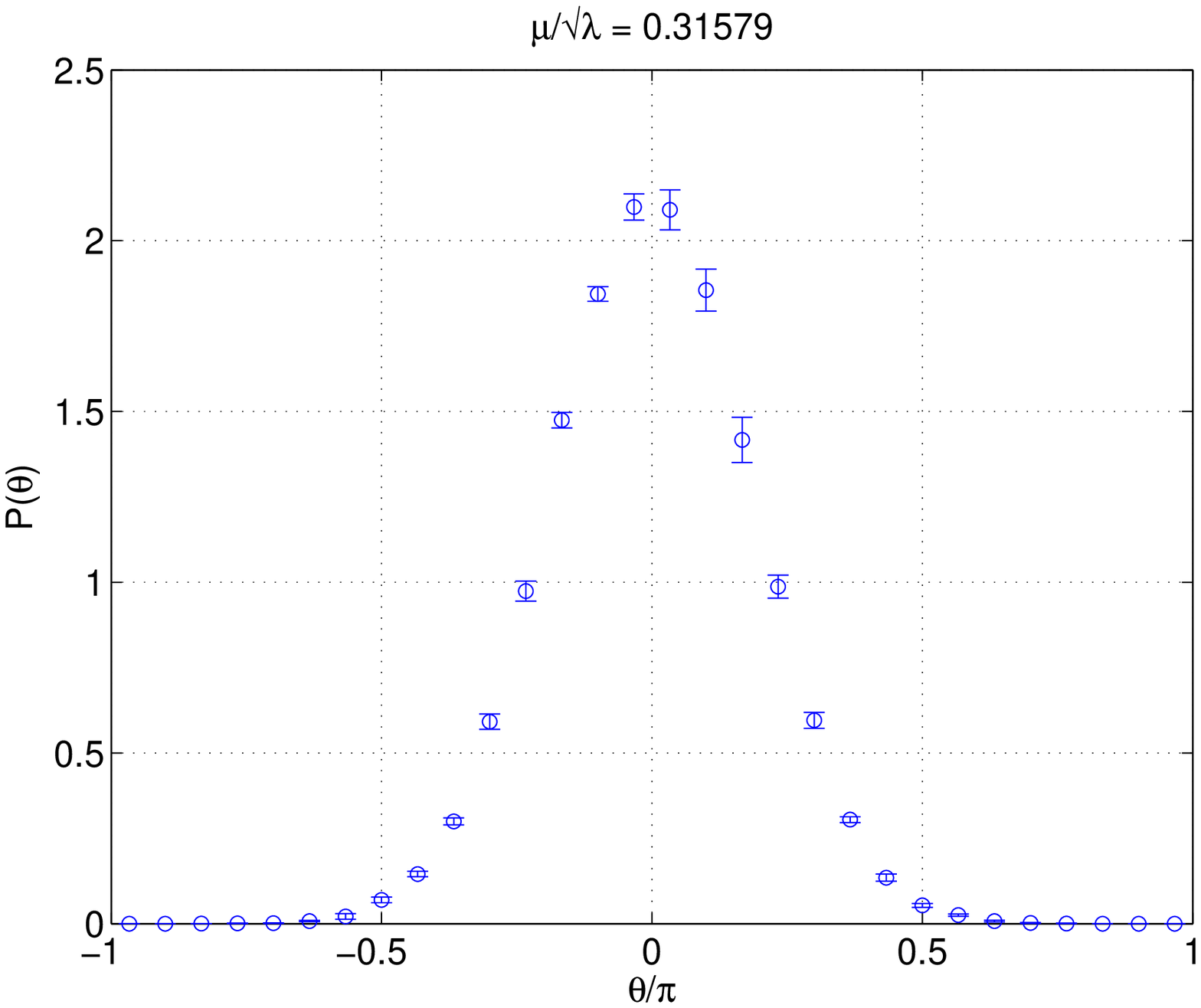}
\includegraphics[width=10cm,height=9.25cm]{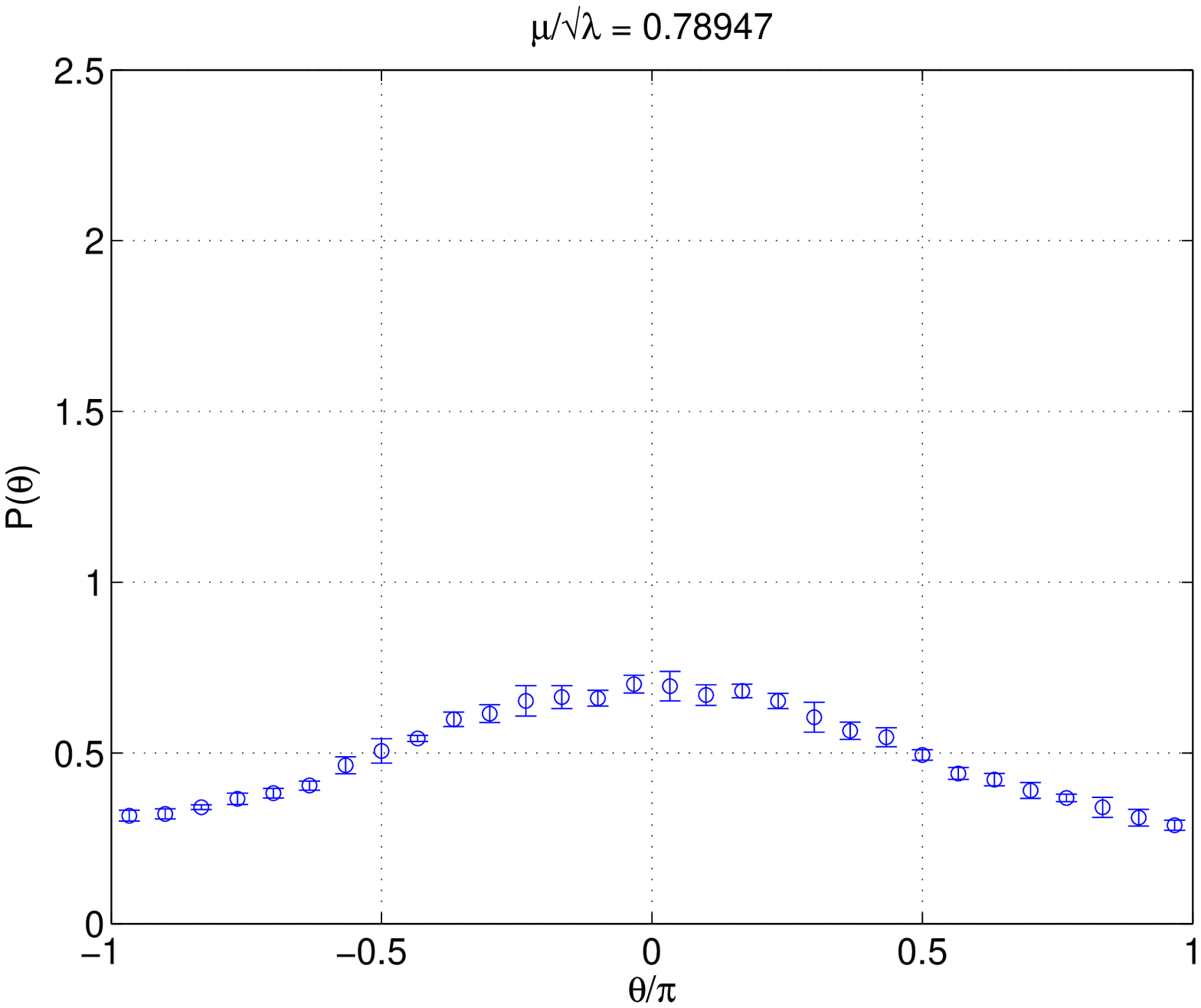}
}
\vspace{1cm}
\centerline{
\includegraphics[width=10cm,height=9.25cm]{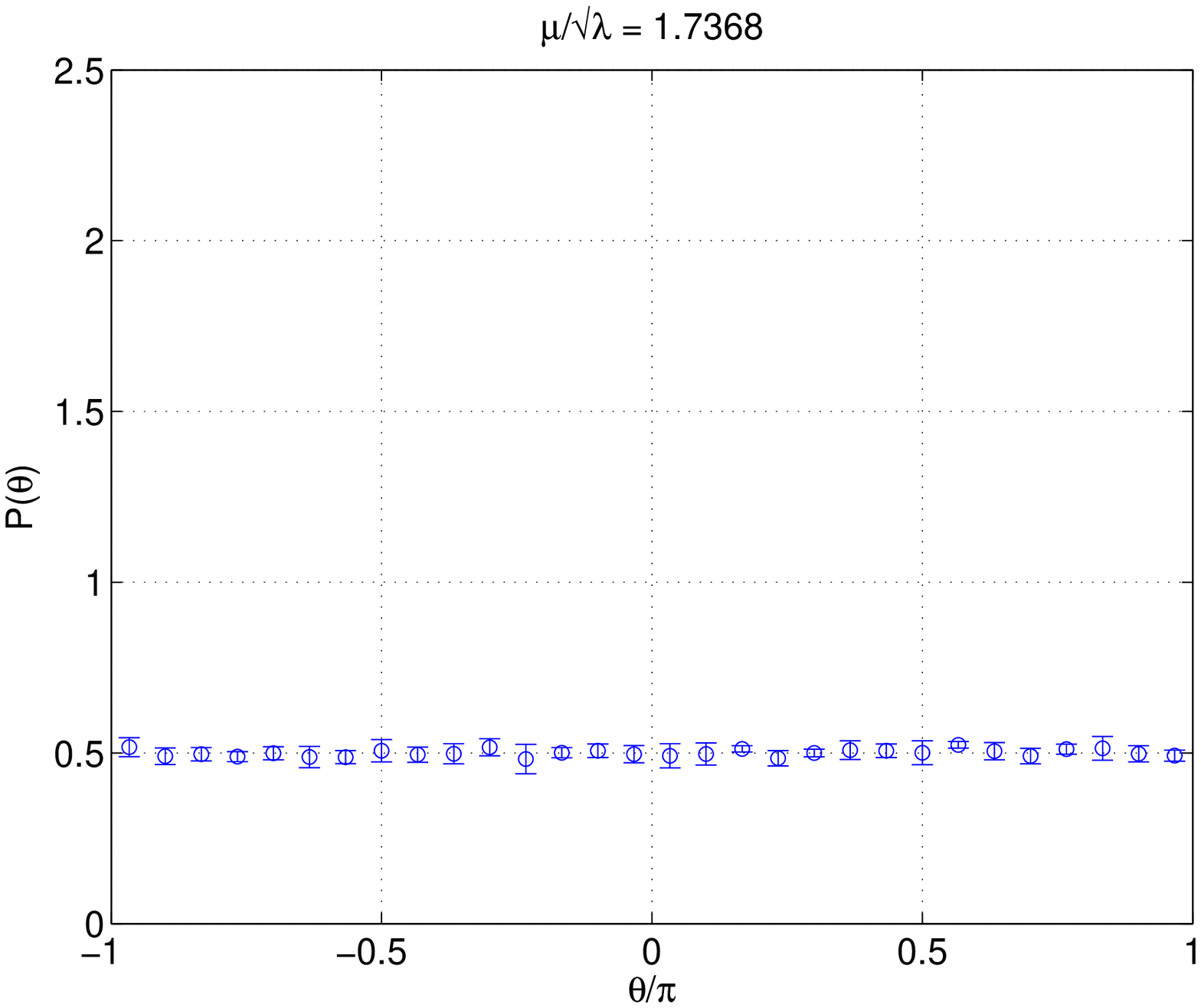}
\includegraphics[width=10cm,height=9.25cm]{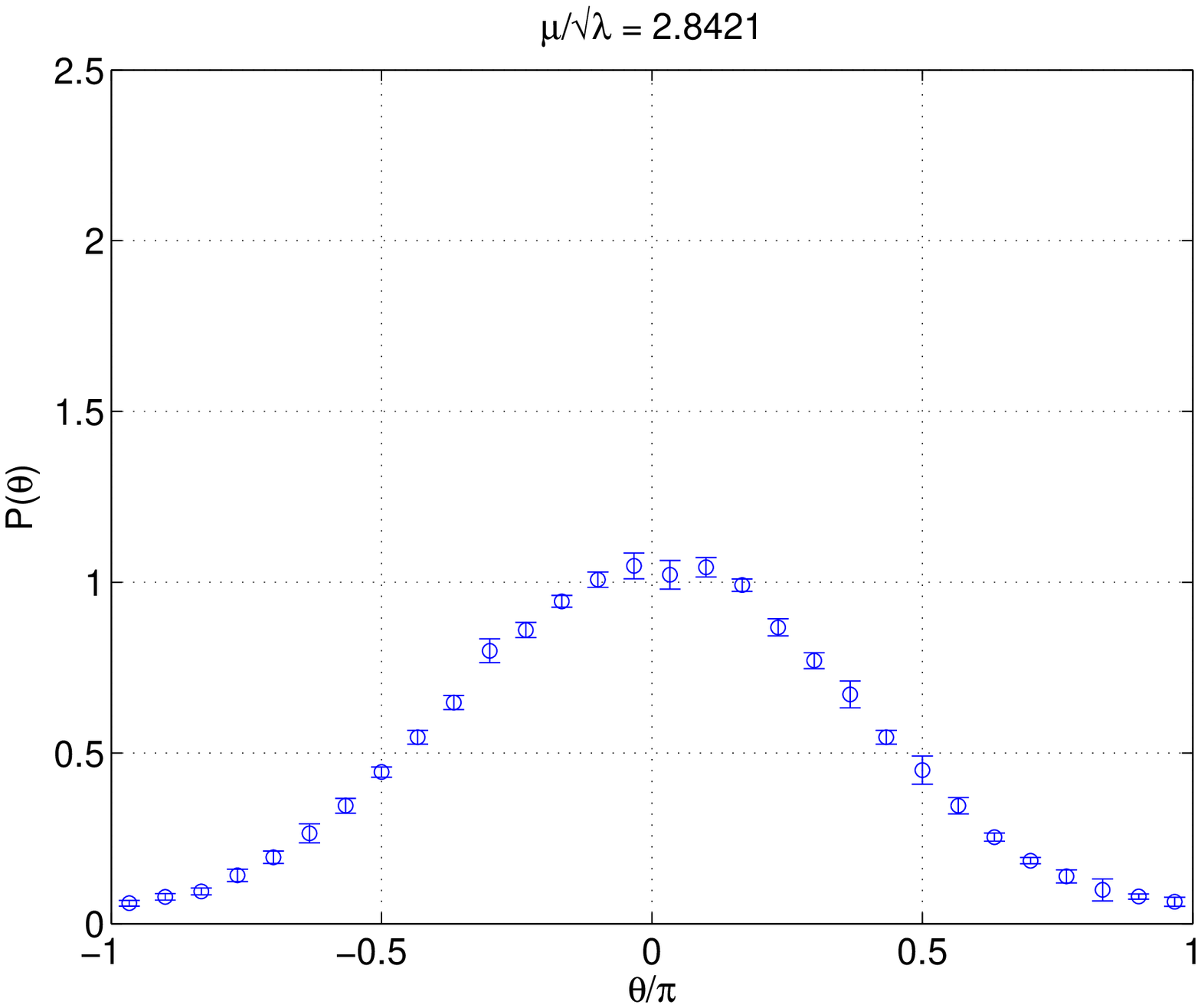}
}
\caption{The probability density of $\theta$ for $SU(20)$, $b=6.0$ and various values of $\mu$. The solid curves are the Gaussian fits (see Table~\ref{fit_res}).
}
\label{Ptheta1}
\end{figure}

The probability distribution of  $\theta$ itself is shown in Fig.~\ref{Ptheta1} where we see that, as \Ref{LSV} anticipates, it is a periodic Gaussian for low values of $\mu$. At larger values of $\mu$ it grows wider and flattens. 
We fitted $P(\theta)$ with the forms
\begin{eqnarray}
P_G(\theta) &\sim & \sum_{n=-1000}^{1000}\, \exp\left[-\left( \theta + 2\pi n\right)^2/\Delta\right]\qquad ({\rm periodic\,\,Gaussian}), \label{Pg}\\
P_L(\theta) &\sim & 1/\left[\Delta - \cos \theta\right]\hskip 3.5cm ({\rm periodic\,\,Lorentzian}).\label{Pl}
\end{eqnarray}
We present the resulting parameters of the fits in Table~\ref{fit_res}. The statistical error on the fit parameters represents the one-sigma error from the fit, as well as the differences between fits done for histograms generated by cutting the full statistical sample into $N_{\rm bin}=15,25$ and $35$ bins. The different values of the $\chi^2$ in the table represent this dependence on $N_{\rm bin}$. Our fits were uncorrelated, and changing $N_{\rm bin}$ is the way we check for the correlation of the different bins in the histogram.
\begin{table}[h]
\setlength{\tabcolsep}{4mm}
\begin{tabular}{|cc|cc|cc|}
\hline\hline
\multirow{2}{*}{$\mu/\sqrt{\lambda}$} & \multirow{2}{*}{Gauge group} & \multicolumn{2}{c|}{Gaussian fit}&\multicolumn{2}{c|}{Lorentzian fit}\\
\cline{3-6}
& & $\Delta$ & $\chi^2/$dof & $\Delta$ & $\chi^2/$dof\\ \hline
\multirow{2}{*}{$0.31579$}& $SU(20)$ & $0.71(3)$& $1.6-2.7$& \multicolumn{2}{c|}{No fit}\\
& $SU(40)$ & $1.17(1)$ & $0.8-1.6$ & \multicolumn{2}{c|}{No fit} \\ \hline
\multirow{2}{*}{$0.78947$}& $SU(20)$ & $6.5(1)$ & $1.8-2.5$& \multicolumn{2}{c|}{No fit}\\
& $SU(40)$ & $11.5(5)$ &$0.9-1.3$& $9.0(7)$ & $0.7-1.0$\\ \hline
\multirow{2}{*}{$1.7368$}& $SU(20)$ & $17(2)$ &  $1.8-3.1$& $40(20)$& $1.8-3.0$\\
& $SU(40)$ & $18(3)$ &$1.2-1.5$ &  $50(30)$ & $1.6-1.5$\\ \hline
\multirow{2}{*}{$2.8421$}& $SU(20)$ & $2.84(4)$ &  $1.0-2.8$  & \multicolumn{2}{c|}{No fit}\\
& $SU(40)$ & $4.65(5)$ & $0.6-0.9$& \multicolumn{2}{c|}{No fit}\\ \hline
\end{tabular}
\caption{Results of fits to the histograms in $\theta$ (see text).
}
\label{fit_res}
\end{table}

From the table we see that while the periodic Gaussian can be fitted to the data for low and high values of $\mu$, the Lorentzian form cannot. At $\mu\simeq 1.7 \surd\lambda$ the Lorentzian becomes consistent with the data, but at that point the histogram is so flat that both the Gaussian and the Lorentzian can be good fits. Note also that the values of $\chi^2$ are not too small, but looking at the plots they seem to reflect a general scatter of the data around the curve and so are likely to come from an underestimate of the statistical error. 

\begin{figure}[h!]
\centerline{
\includegraphics[width=10cm,height=7cm]{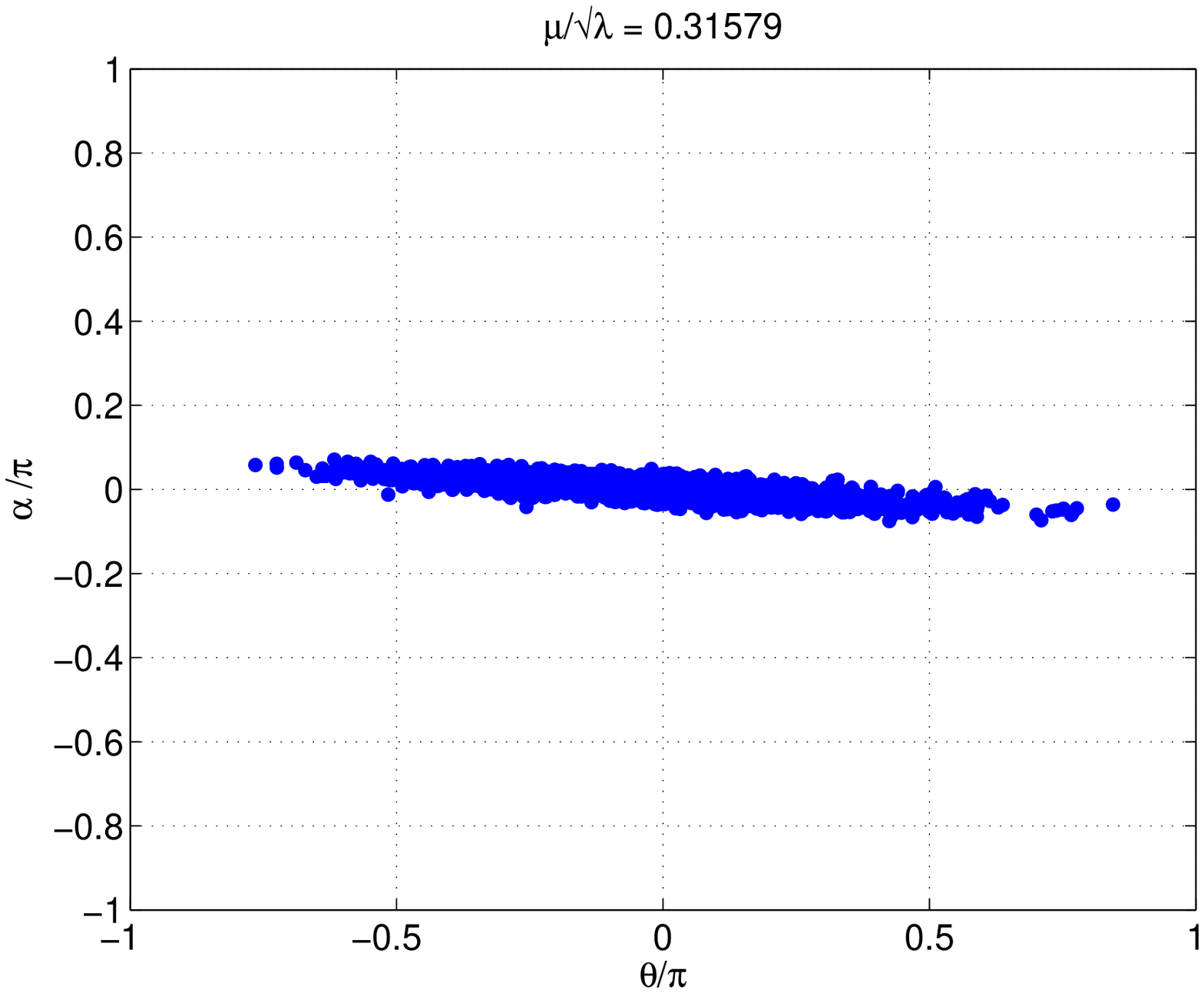}
\includegraphics[width=10cm,height=7cm]{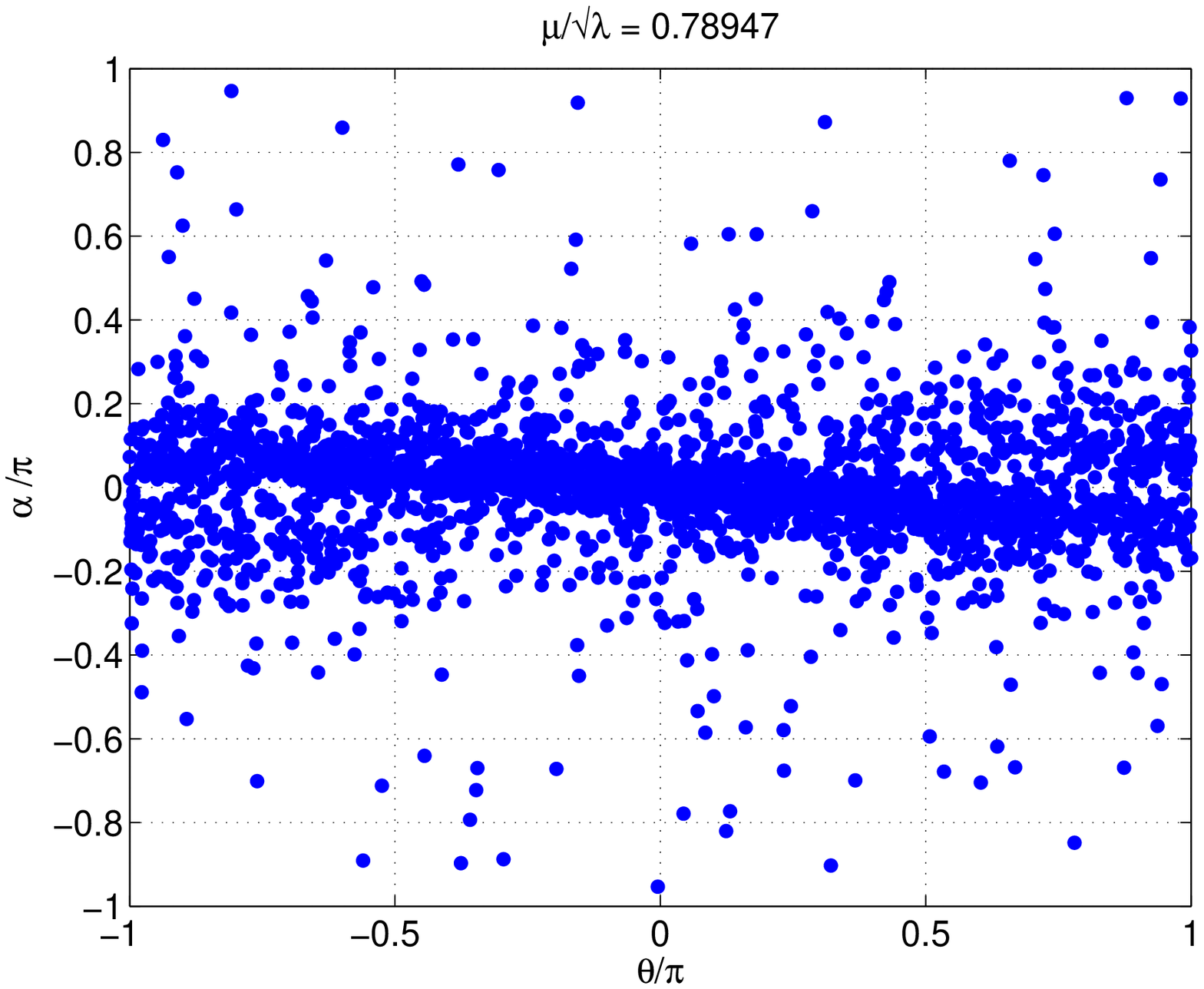}
}
\centerline{
\includegraphics[width=10cm,height=7cm]{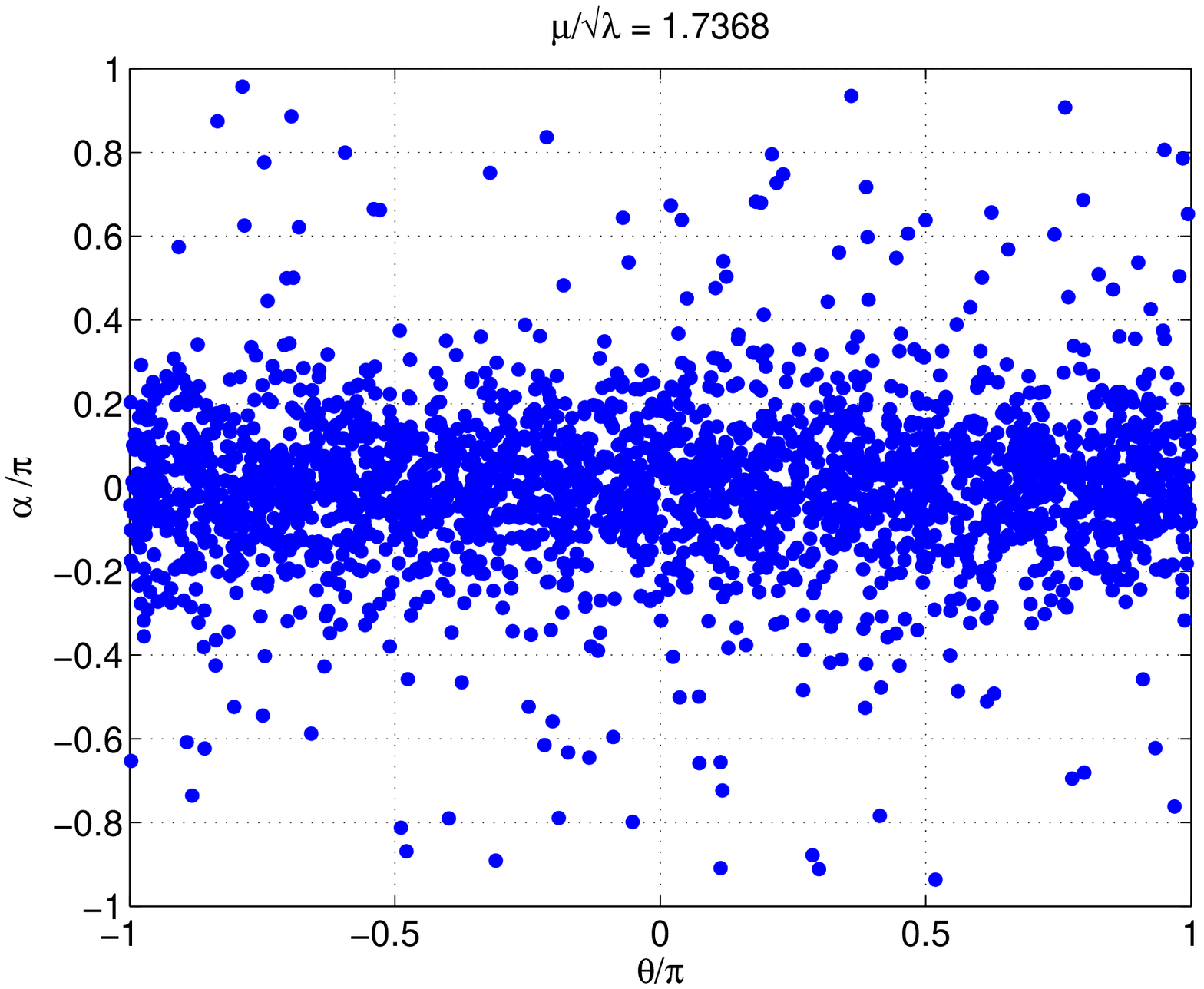}
\includegraphics[width=10cm,height=7cm]{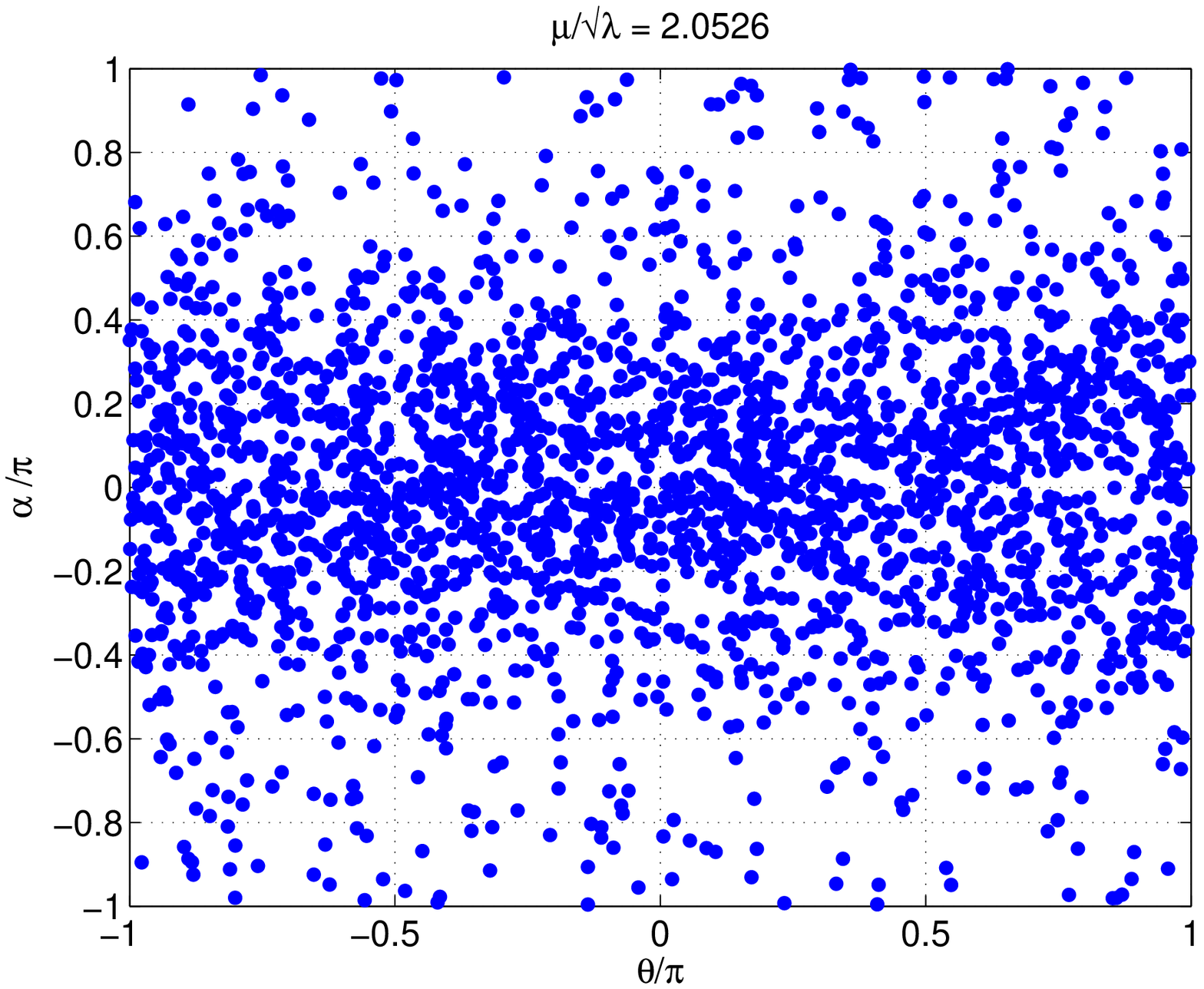}
}
\centerline{
\includegraphics[width=10cm,height=7cm]{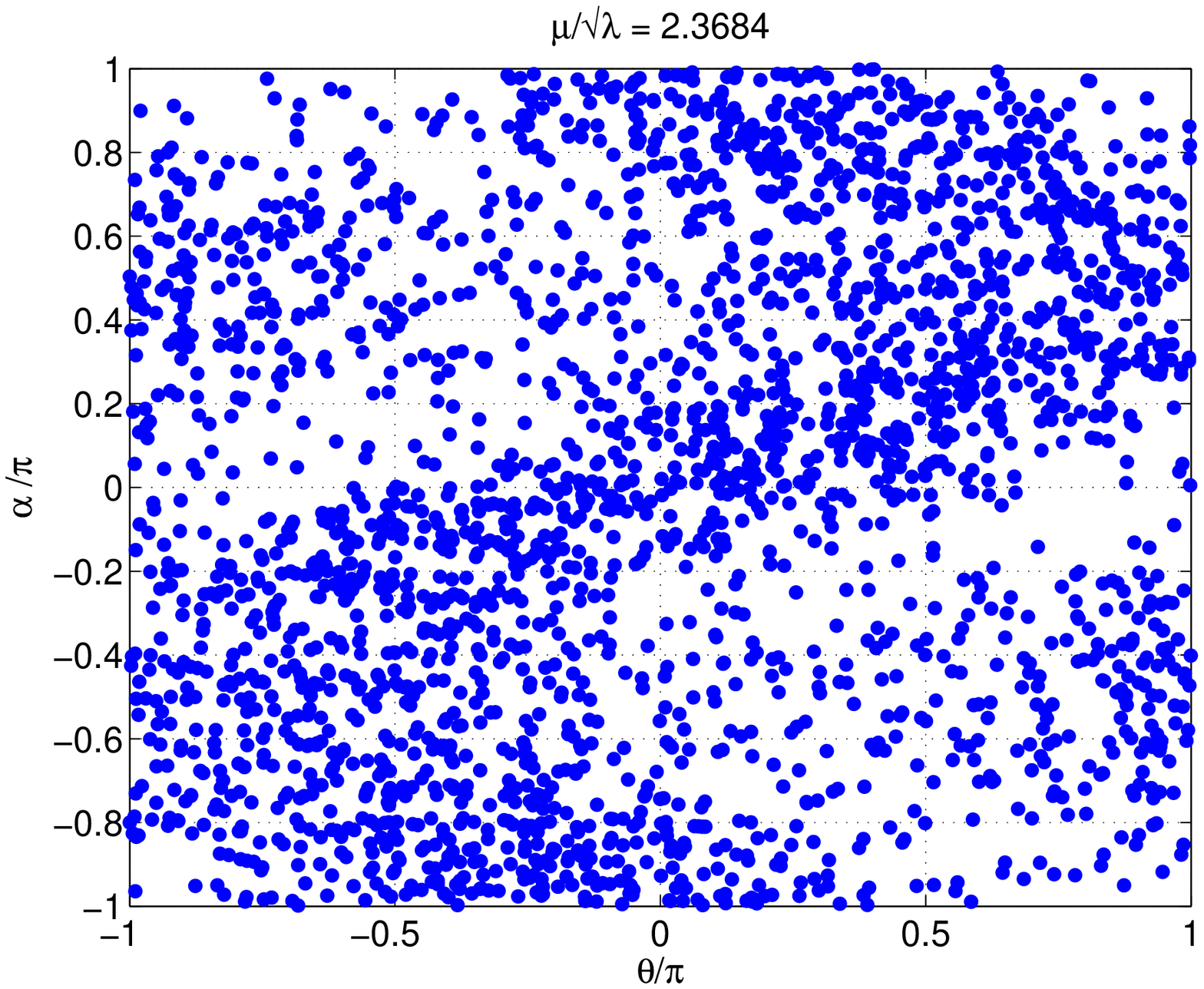}
\includegraphics[width=10cm,height=7cm]{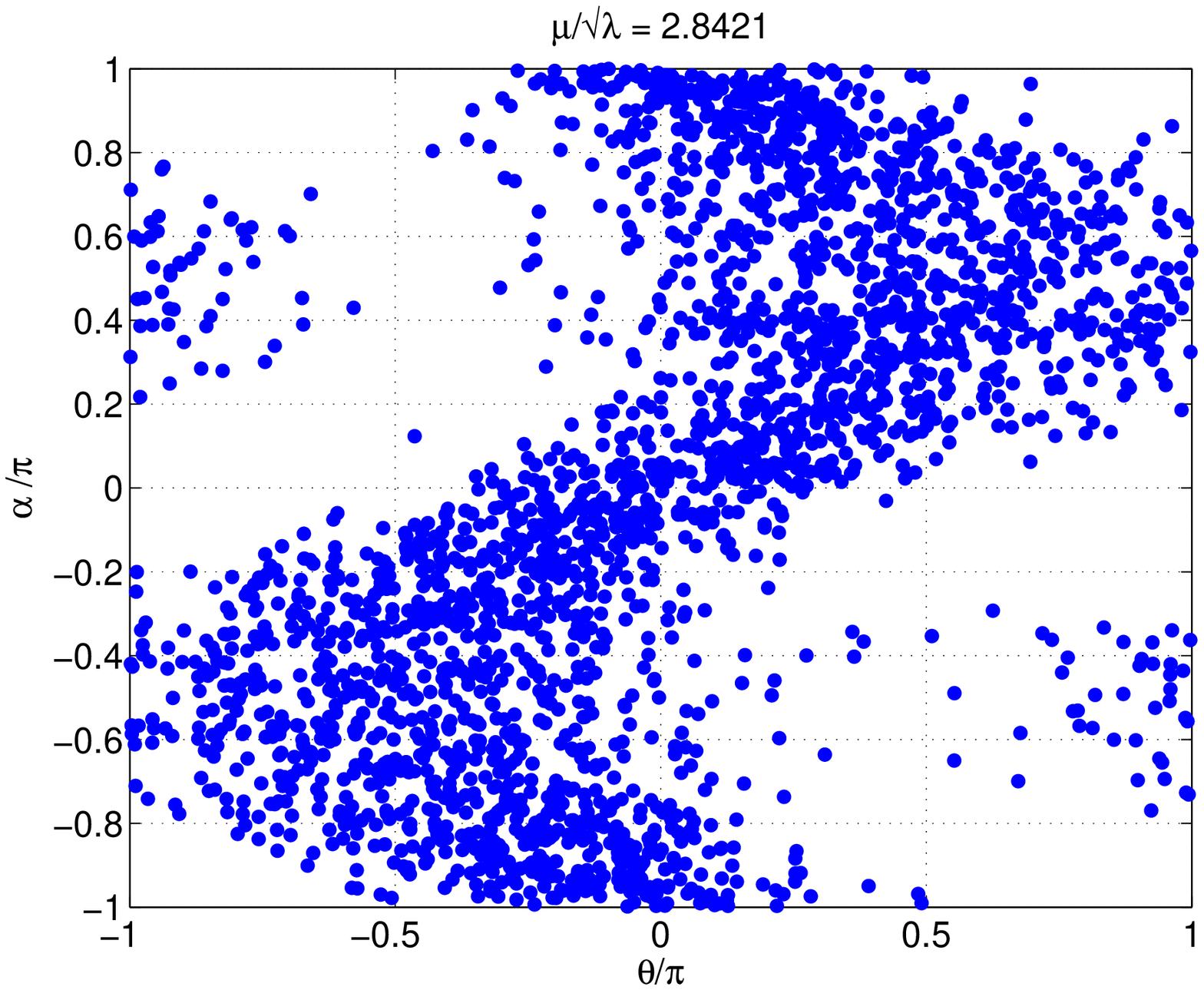}
}
\caption{Same as Fig.~\ref{corr1}, but for $SU(40)$.
}
\label{corr2}
\end{figure}

\begin{figure}[t]
\centerline{
\includegraphics[width=10cm,height=8cm]{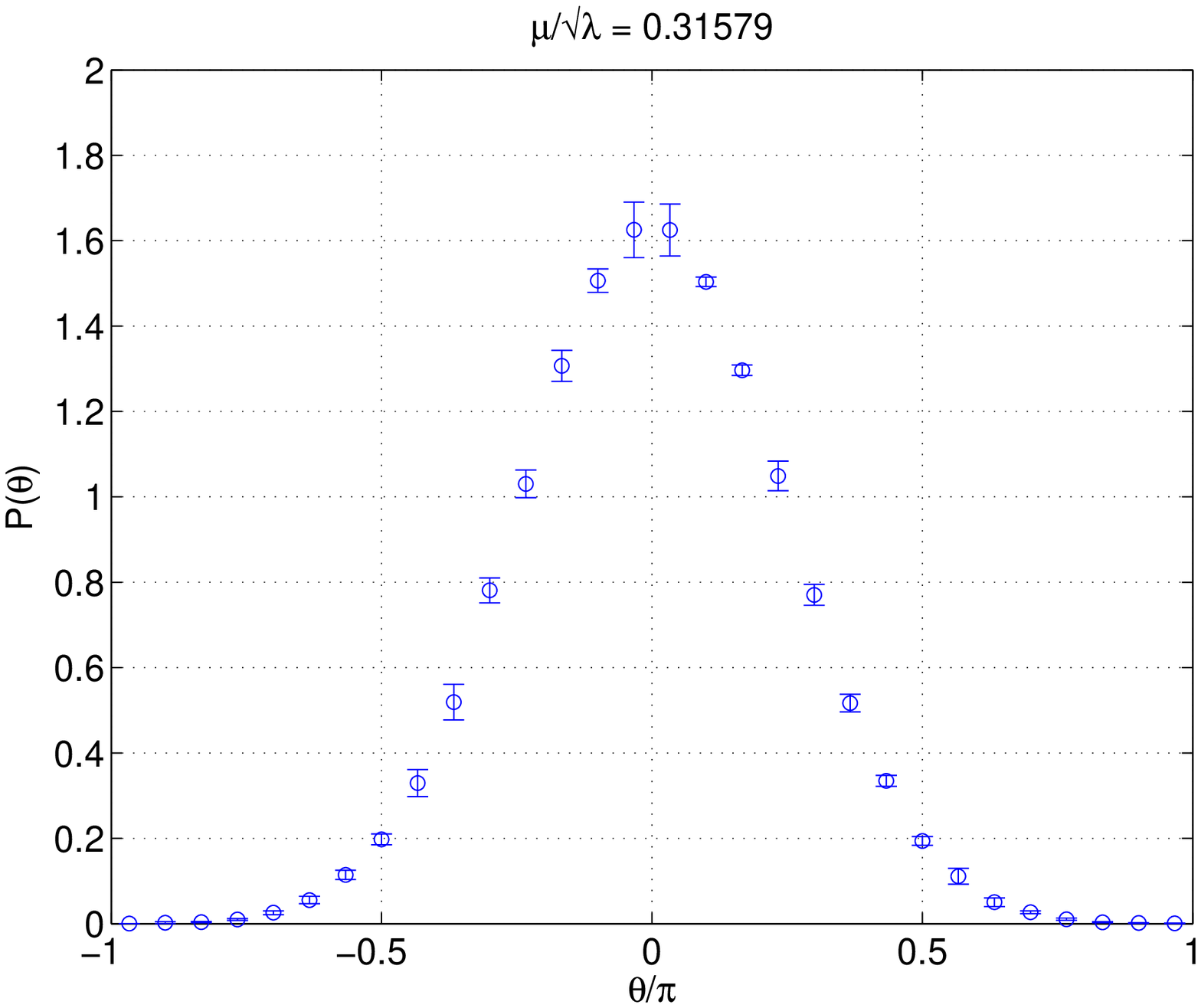}
\includegraphics[width=10cm,height=8cm]{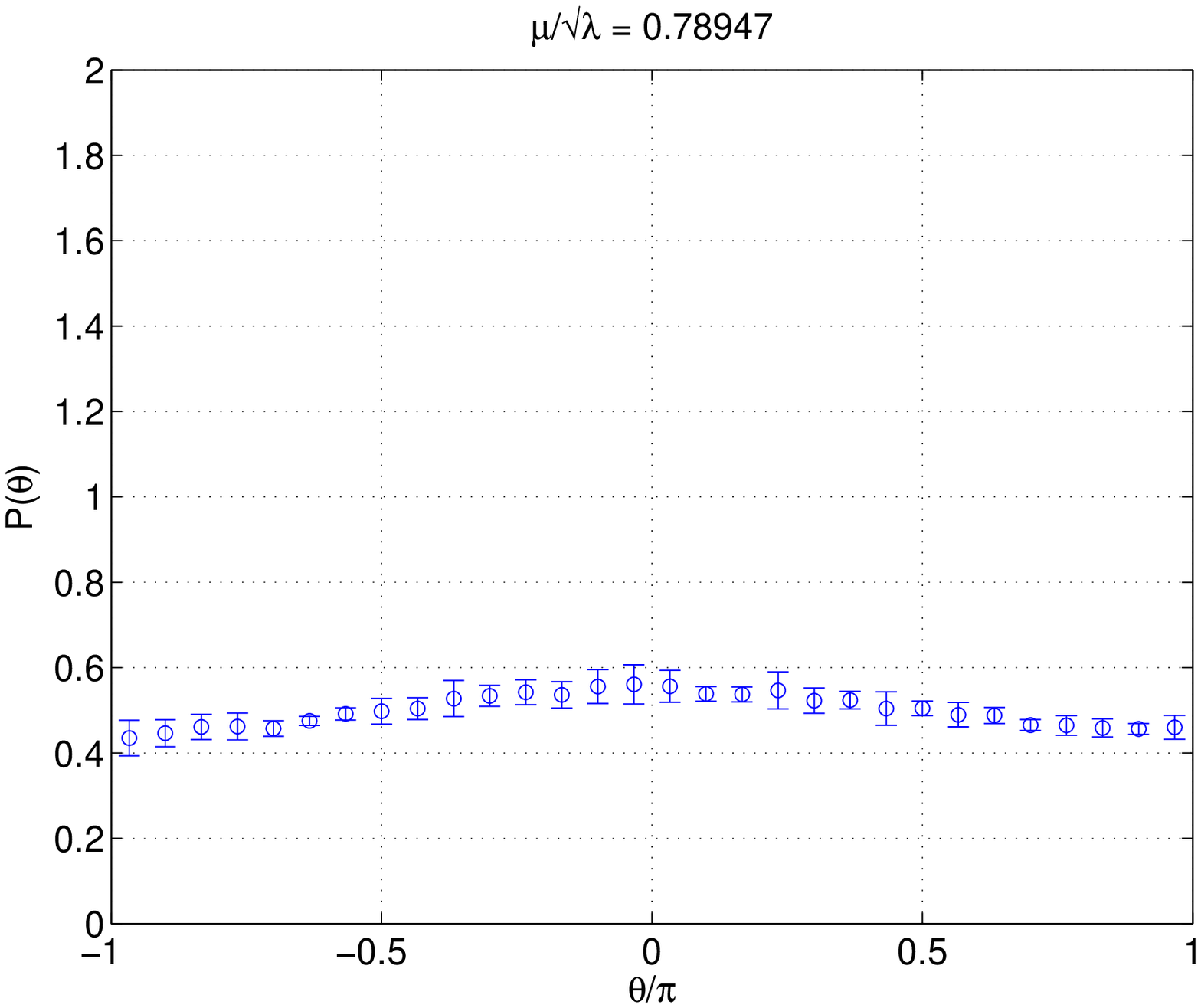}
}
\centerline{
\includegraphics[width=10cm,height=8cm]{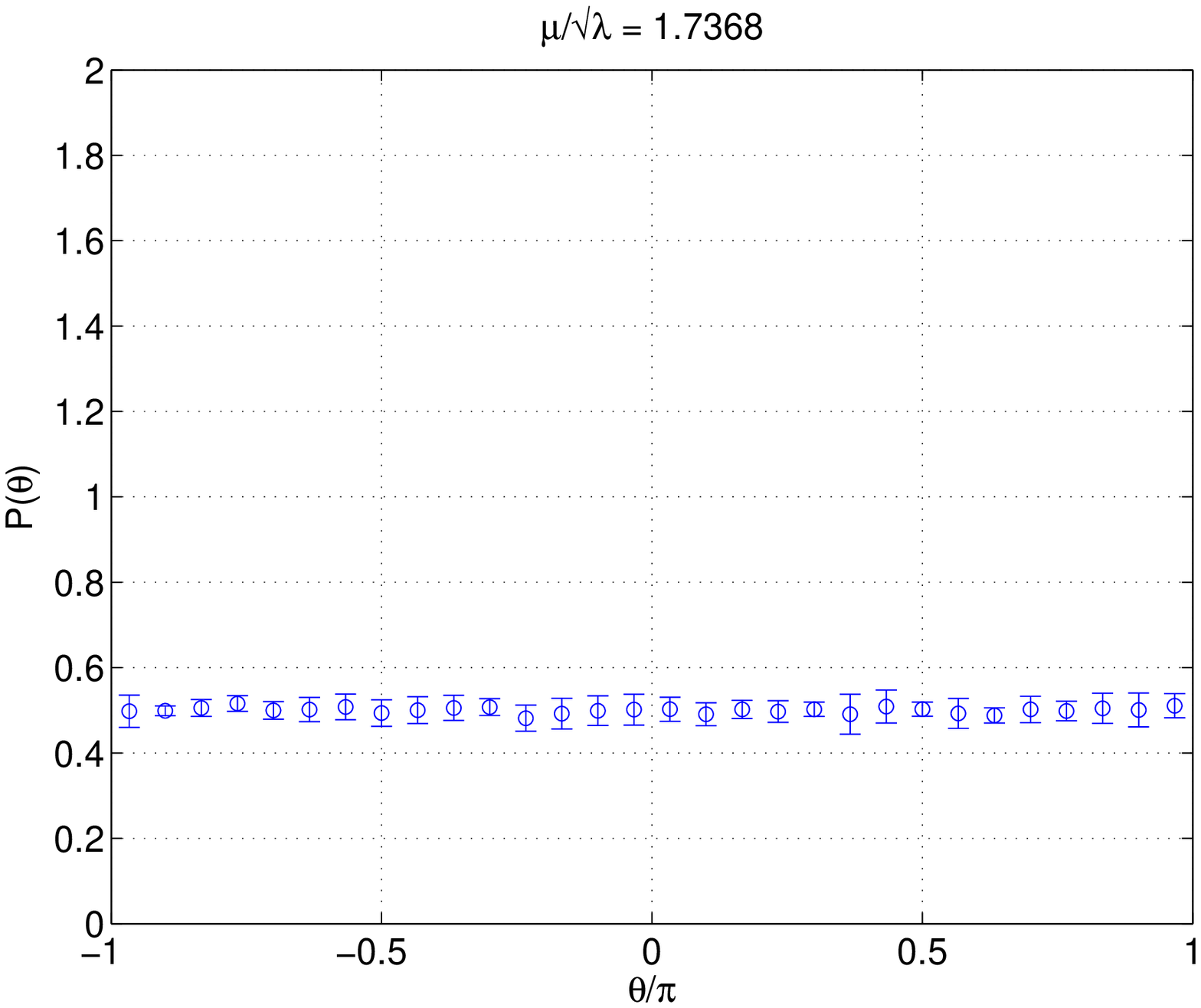}
\includegraphics[width=10cm,height=8cm]{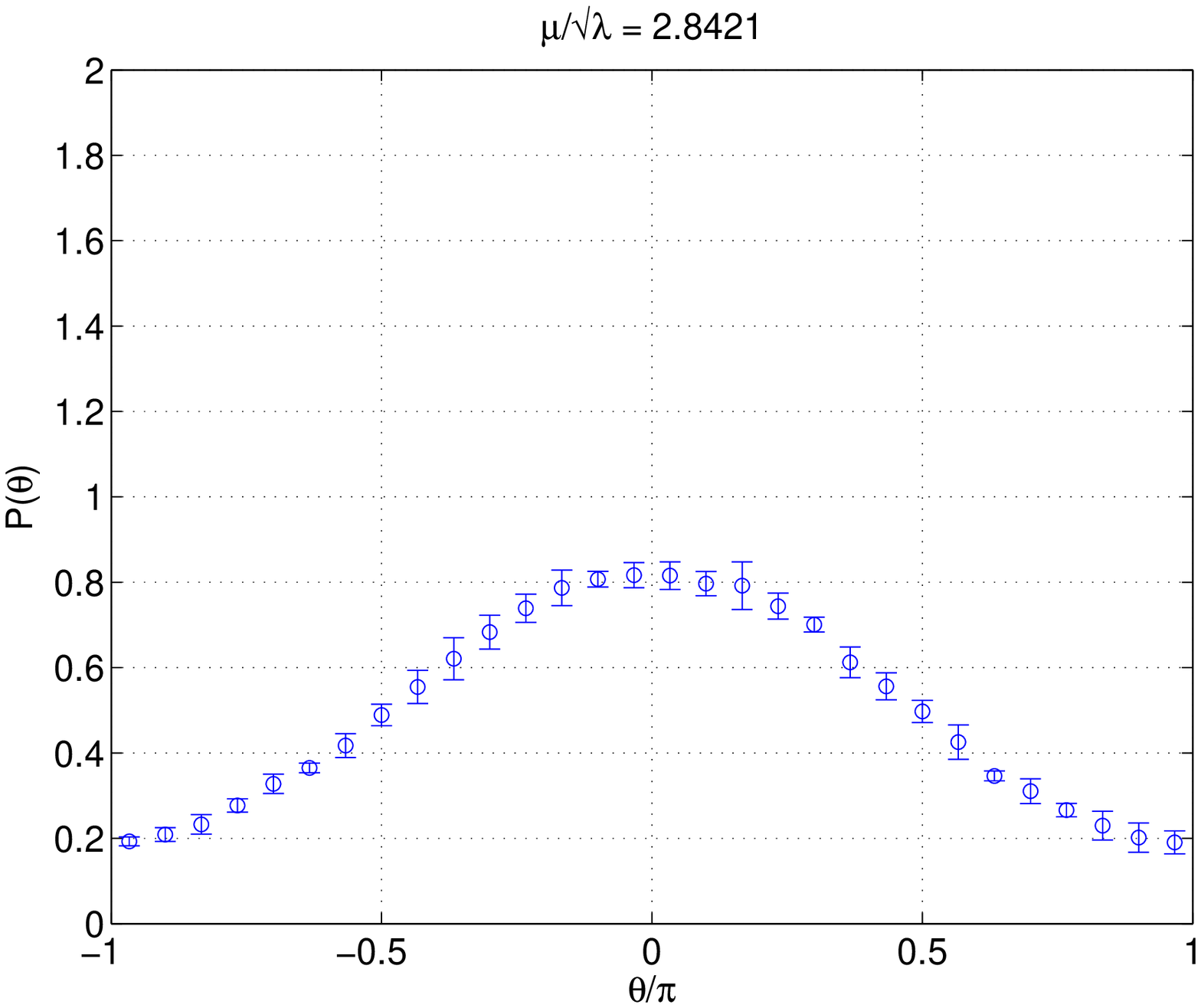}
}
\caption{The same as Fig.~\ref{Ptheta1}, but for $SU(40)$. The only panel that includes both the Gaussian and the Lorentzian fit is that of $\mu\simeq 0.79 \surd\lambda$. The Gaussian curve is a solid line (blue), and the Lorentzian is the dashed (red). It is clear that both are very close to each other.}
\label{Ptheta2}
\end{figure}

Let us now ask how Fig.~\ref{corr1} and Fig.~\ref{Ptheta1} change when we increase $N$. The results for $SU(40)$ are given in Figs.~\ref{corr2}-\ref{Ptheta2} and we see that increasing $N$ makes the distribution of both $\theta$ and $\alpha$ more spread out.
 The corresponding fitting results to the distribution $P(\theta)$ in this case are also presented in the Table~\ref{fit_res}.

It is hard to conclude from the figures how the correlations between $\theta$ and $\alpha$ changes with $N$. Nevertheless, the widening of the histograms in Fig.~\ref{Ptheta2} compared to Fig.~\ref{Ptheta1} is qualitatively consistent with \Ref{LSV} if we interpret the large-$N$ limit as the thermodynamical limit. More accurate studies are required to pinpoint the nature of the correlation between $\bar\psi\psi$ and $\det D$ (the absolute values of $\bar\psi\psi$ and $\det D$ may play a role also). Probably the best indicator for this is the connected correlation function between these operators, which is nothing but the difference between the unquenched $\<\bar\psi\psi\>$ and $\<\bar\psi\psi\>_{\rm quenched}$. As we mentioned in \Sec{physical},  this difference is significantly nonzero and independent of $N$ for $\mu>m_\pi/2$. 

We note in passing that we see that the $\chi^2$ for $SU(40)$ cases are lower than for $SU(20)$ and this may reflect the presence of significant $1/N$ corrections that are required to make the the fitting ansatze in Eqs.~\ref{Pg}--\ref{Pl} consistent with our data. Also, we see that, as predicted by chiral perturbation theory, the Lorentzian is consistent with our data at $\mu\simeq 0.8\surd\lambda$. This should only be seen as partial evidence to support these predictions since for the same values of $\mu$, the Gaussian is also consistent with our data, and because for $SU(20)$ it is only the Gaussian that provides an acceptable fit at that value of $\mu$.  Finally we note the way $\Delta$ depends on $N$ is expected to be linear for the Gaussian case (see \Ref{LSV}, but this is only partially consistent with what we see in Table~\ref{fit_res}; this is  expected to happen only for $\mu\simeq 0.32 \surd\lambda<m_\pi/2$, but there we see that $\Delta_{SU(40)}/\Delta_{SU(20)}=1.65(7)$ instead of $2$. In the case of $\mu\simeq 2.8\surd\lambda$, where $P(\theta)$ is clearly Gaussian and not a Lorentzian, we see that $\Delta_{SU(40)}/\Delta_{SU(20)}=1.64(3)$. We do not fully understand these results which actually correspond to $\Delta_{SU(N)} \sim N^{0.71(3)-0.72(6)}$. They may reflect significant $1/N$ corrections in the $SU(20)$ case.

\begin{figure}[t]
\centerline{
\includegraphics[width=12cm,angle=-90]{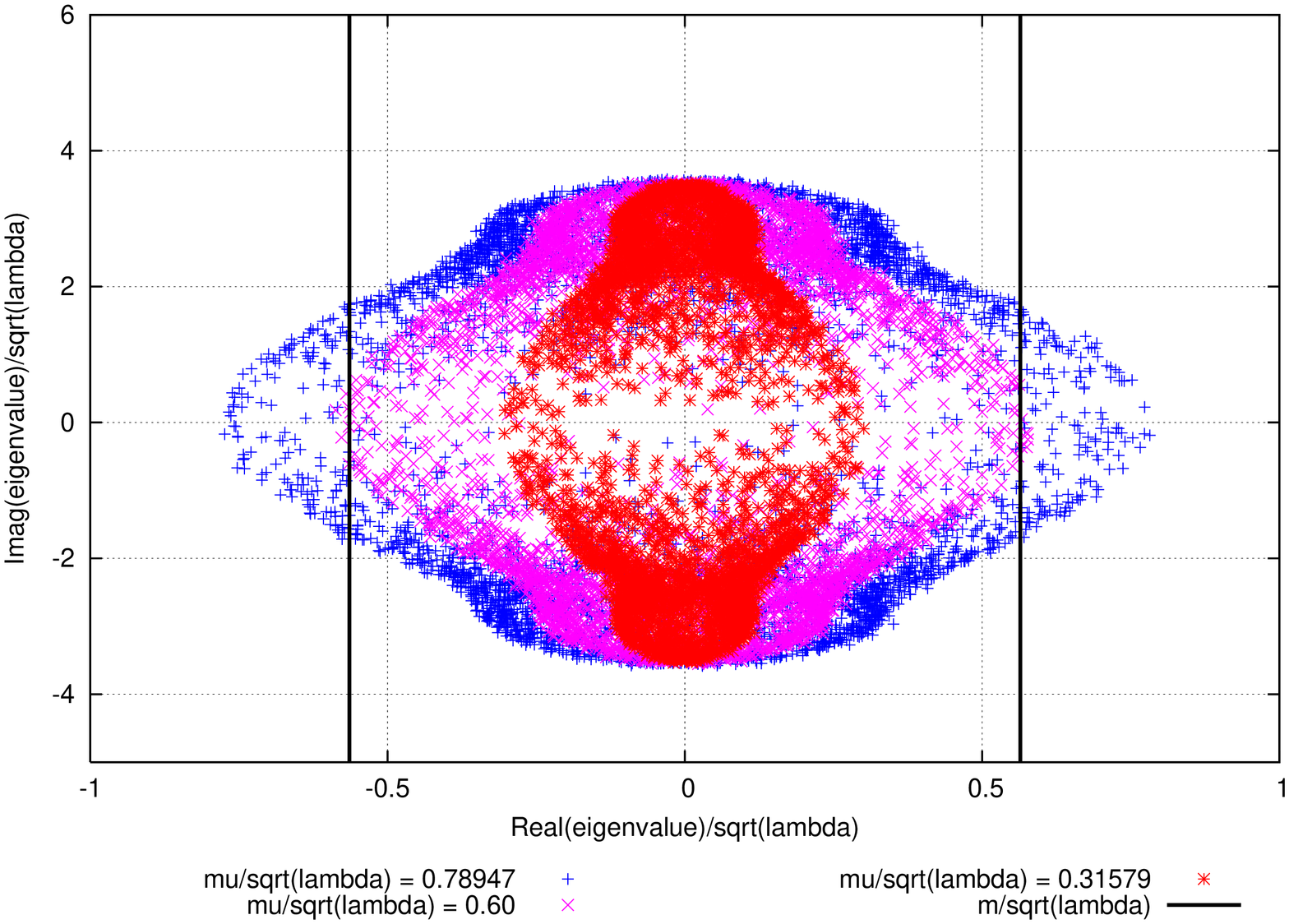}
}
\caption{The scatter in the complex plane of the eigenvalues of $D$. Here $b=6.00$ and the gauge group was $SU(40)$.
}
\label{eigsct}
\end{figure}

We end this section by presenting the way the eigenvalues of $D$ are scattered in the complex plane. According to chiral perturbation theory, when $\mu<m_\pi/2$, the bare quark mass $m$ is outside the support of the eigenvalue distribution, while for $\mu>m_\pi/2$ it is within it. This is confirmed in Fig.~\ref{eigsct} where we show the eigenvalue scatter of $50$ gauge configurations for $SU(40)$ and $b=6.00$. We present the cases of $\mu/\sqrt{\lambda}=0.31579, 0.60, 0.78947$ which are below, close to, and above $\frac12 m_\pi/\sqrt{\lambda}$. Also, the cases of $\mu/\sqrt{\lambda}\simeq 0.31579, 0.78947$ correspond to the two upper panels of Figs.~\ref{corr2}--\ref{Ptheta2}. Recall that in our current work we fix the quark mass to be  $m/\sqrt{\lambda}=1/\sqrt{\pi}\simeq 0.5642$ (see \Eq{gamma}).

\section{Summary and conclusions}
\label{summary}

In this paper we studied the sign problem of the euclidean path integral of large-$N$ QCD at nonzero chemical potential $\mu$. To do so we used the Eguchi-Kawai (EK) equivalence which allows us to approach the problem via the volume reduced version of theory. For reasons of computational cost we focused on the two dimensional case in this paper. Extensions to four dimensions are straight forward and we discuss them below. 

We used lattice Monte-Carlo simulations to explore the sign problem numerically and did so for three lattice spacing, and for $SU(N)$ gauge groups with $N\le 60$. We measured the average sign of $\det D$, and saw that it decreases to zero when  $\mu\simeq m_\pi/2$, but then rises back to $1$ when the lattice site is saturated with quarks.

We proposed a way to suppress the sign fluctuations in the hadronic phase which amounts to replacing the pure-gauge average of $\det D$, by an average over a real and positive quantity. We denote the latter by $\left(\det D\right)_Z$, and it is the average of $\det D$ over a set of gauge configurations that are related by $Z_N$ center transformations. We call this method `$Z_N$-averaging' (see also \cite{early}) and we test it numerically. We find it to be successful and that, in the large-$N$ volume-reduced system, the average sign of $\left(\det D\right)_Z$ is $1$ for a wide range of $\mu$ that includes all the hadronic phase. Thus, this method removes the sign fluctuations from the hadronic phase. The computational cost of $Z_N$ averaging grows linearly in $N$, and this means it provides an exponential gain in calculating the average of $\det D$, since the computational cost of overcoming the sign fluctuations involved in such an average by brute force grows exponentially in $N$. This means that $Z_N$ averaging solves the silver blaze problem in our volume-reduced large-$N$ theory.

$Z_N$-averaging also identifies the regime in $\mu$ where the true, non-silver-blaze, sign problem of the large-$N$ reduced theory is most severe. The latter  happens beyond the hadronic phase, just before the saturation regime. We identified the saturation regime also in measurements of the free energy, which becomes linear in $\mu$ in that regime. Our measurements of the free energy and of the chiral condensate are unquenched and as anticipated physically, we see that they show $\mu$-independence throughout the hadronic phase. Our measurements of the {\em quenched} chiral condensate serve to contrast this since they show that $\<\bar\psi\psi\>_{\rm quenched}$ changes at around $\mu\simeq m_\pi/2$, where nothing happens to the free energy and to the physical unquenched condensate. This reflects the breakdown of the quenched approximation.

In an attempt to numerically understand the failure of the quenched approximation mentioned above, we analyzed the way the operators $\bar\psi\psi$ and $\det D$ are correlated with each other for different values $\mu$. Specifically, we checked whether their correlations is mostly due to a correlation between the phases of the operators, but were not able to unambiguously conclude if this is the case or not. Instead, we directly calculated the connected correlation of $\bar\psi\psi$ and $\det D$. This correlation is nothing but the difference between the quenched and unquenched chiral condensates. We saw that this difference becomes large for $\mu\stackrel{>}{_\sim}m_\pi/2$, and that it does not go down with increasing $N$. This means that the failure of the quenched prescription can be thought of as the breakdown of large-$N$ factorization. It would be useful to understand this breakdown from a physical point of view rather than a numerical one. It will also be useful to understand our results for the difference between the quenched and unquenched condensates at $\mu=0$ and verify that they go away at $N\to \infty$ by performing simulations for large values of $N$.

We also measured the way the phase of $\det D$ is distributed as a function of $N$ and $\mu$, and saw that, in agreement with  chiral perturbation theory (see \Ref{LSV}), these distributions are Gaussian for $\mu<m_\pi/2$ and become extended when $\mu$ increases beyond that. The width of these distributions increases towards the thermodynamical limit of $N\to\infty$. At the quantitative level, our results support the predictions of \Ref{LSV} only partially, but this may be due to $1/N$ effects. Finally, we confirmed that when $\mu>m_\pi/2$ the quark mass $m$ enters the support of the density of eigenvalues of $D$. 

The analysis we presented here can in principle be repeated for four dimensions but this will require more work. In particular, in $4d$, the straight forward EK theory is not equivalent to large-$N$ QCD (even in the hadronic phase) and other prescriptions, like those suggested in Refs \cite{DEK} and \cite{AEK} (and studied numerically in Refs.~\cite{BS} and \cite{HN}), are needed to overcome this issue. Once this is done, and one has a set of pure-gauge field configurations in hand, then the methods of this paper can be used with a modest increase in computational effort. Specifically, the only change is that the dimension of the Dirac matrix $D$ is doubled relative to the two-dimensional case. We therefore estimate that future $4d$ studies of the issues we explore in this work are feasible. 

Finally, we proposed that $Z_N$ averaging solves the silver blaze problem when the contribution to the Dirac operator determinant from baryonic worldlines (or Polyakov loops whose winding number is a multiple of $N$) is suppressed on a configuration-by-configuration basis (for an explanation of this see \Sec{2scenarios}).  In the current paper we see numerically that in $1+1$ dimensions such suppression indeed takes place in the large-$N$ volume-reduced version of the QCD, but we do not know if this is also the case for physical (four dimensional and $3$-color) QCD. More precisely, in \Sec{2scenarios} we discussed how such a suppression will {\em not} happen at $T\to 0$ since, in that limit, a configuration with $k$ baryonic worldlines will most likely be only moderately suppressed by the Boltzmann factor of $e^{-kNm_\pi/(2T)}$. We argued, however, that there may be a region of $T>0$ where, instead, these worldlines might  be suppressed  and where $Z_N$ averaging would work. This `golden region' is the analog of the `golden window' in the studies of baryon correlation functions \cite{Detmold}, where one searches for optimized baryonic wave functions whose noise couples very little to pion systems. Therefore it seems interesting to attempt $Z_N$ averaging in physical QCD (with light quarks) and try and locate this golden region.

\section*{Acknowledgments}

For useful discussions and correspondences I thank A.~Alexandru, S.~Chandrasekharan, Ph.~d.~Forcrand, R.~Narayanan, K.~F.~Liu, Weonjong Lee, M.~P.~Lombardo, S.~R.~Sharpe, K.~Splittorff, J.~Verbaarschot, and L.~G.~Yaffe. I wish to thank G.~Aarts and S.~Chandrasekharan for organizing the interesting  ``Sign Problems and Complex Actions'' workshop at the ECT*, the ECT* for its kind hospitality, and the KITP-Beijing, where part of this study was completed, for its support. 
Finally, for constant encouragement and support I thank Sarit Goren. 
This work was supported by the U.S. Department of Energy under Grant No. DE-FG02-96ER40956 and also in part by the Project of Knowledge Innovation Program (PKIP) of Chinese Academy of Sciences, Grant No. KJCX2.YW.W10.

\end{document}